\documentclass[aip,jap,reprint,10pt,showpacs,altaffillsymbol,superscriptaddress,longbibliography]{revtex4-1}
\pdfoutput=1
\usepackage{graphicx}
\usepackage{amsfonts}
\usepackage{amsmath}
\usepackage{natbib}
\usepackage{color}

\begin{document}

\title{Realisation of a quantum current standard at liquid helium temperature with sub-ppm reproducibility}

\author{Stephen~P.~Giblin}
\affiliation{National Physical Laboratory, Hampton Road, Teddington, Middlesex TW11 0LW, United Kingdom}
\author{Emma Mykk\"anen}
\affiliation{VTT Technical Research Centre of Finland Ltd, National Metrology Institute VTT MIKES, Espoo, FI-02044 VTT, Finland}
\author{Antti Kemppinen}
\affiliation{VTT Technical Research Centre of Finland Ltd, National Metrology Institute VTT MIKES, Espoo, FI-02044 VTT, Finland}
\author{Pekka Immonen}
\affiliation{VTT Technical Research Centre of Finland Ltd, National Metrology Institute VTT MIKES, Espoo, FI-02044 VTT, Finland}
\author{Antti Manninen}
\affiliation{VTT Technical Research Centre of Finland Ltd, National Metrology Institute VTT MIKES, Espoo, FI-02044 VTT, Finland}
\author{M\'at\'e Jenei}
\affiliation{QCD Labs, QTF Centre of Excellence, Department of Applied Physics, Aalto University, PO Box 13500, FI-00076 AALTO, Finland}
\author{Mikko M\"ott\"onen}
\affiliation{QCD Labs, QTF Centre of Excellence, Department of Applied Physics, Aalto University, PO Box 13500, FI-00076 AALTO, Finland}
\author{Gento Yamahata}
\affiliation{NTT Basic Research Laboratories, NTT Corporation, 3-1 Morinosato Wakamiya, Atsugi, Kanagawa 243-0198, Japan}
\author{Akira Fujiwara}
\affiliation{NTT Basic Research Laboratories, NTT Corporation, 3-1 Morinosato Wakamiya, Atsugi, Kanagawa 243-0198, Japan}
\author{Masaya Kataoka}
\affiliation{National Physical Laboratory, Hampton Road, Teddington, Middlesex TW11 0LW, United Kingdom}

\email[stephen.giblin@npl.co.uk]{Your e-mail address}

\date{\today}

\begin{abstract}
A silicon electron pump operating at the temperature of liquid helium has demonstrated repeatable operation with sub-ppm accuracy. The pump current, approximately $168$~pA, is measured by three laboratories, and the measurements agree with the expected current $ef$ within the uncertainties which range from $0.2$~ppm to $1.3$~ppm. All the measurements are carried out in zero applied magnetic field, and the pump drive signal is a sine wave. The combination of simple operating conditions with high accuracy demonstrates the possibility that an electron pump can operate as a current standard in a National Measurement Institute. We also discuss other practical aspects of using the electron pump as a current standard, such as testing its robustness to changes in the control parameters, and using a rapid tuning procedure to locate the optimal operation point.
\end{abstract}

\pacs{1234}

\maketitle

\section{introduction}

Moving electrons one at a time is a conceptually simple and elegant way to generate an accurate reference current. Following the 2019 re-definition of the International System of Units (SI), such a method is also a most direct way to realise the SI base unit ampere\cite{kaneko2016review,scherer2019single}, requiring only a traceable measurement of the clock frequency $f$. Mesoscopic devices which aim to achieve this controlled electron transport, electron pumps and turnstiles, have been the subject of research for more than 30 years\cite{pekola2013single}. In the last 10 years, pumps based on semiconductor quantum dots have made remarkable progress\cite{kaestner2015non,giblin2019evidence}, and devices based on silicon\cite{zhao2017thermal} and gallium arsenide\cite{stein2016robustness} have demonstrated an accuracy approaching $1$ part in $10^{7}$. The general metrological utility of electron pumps requires, in addition to the absolute accuracy, a broader class of properties to be demonstrated. These include reproducibility of the pump operation across multiple cool-downs, reliability of device fabrication, and operation under conditions which are accessible in a wide range of National Measurement Institutes (NMIs). A simple procedure for tuning the pump is also highly desirable. Thus far, bench-mark experiments such as those reported in Refs. \onlinecite{zhao2017thermal,stein2016robustness} have required sophisticated cryogenic infrastructure; a helium-3 refrigerator\cite{zhao2017thermal} and a dilution refrigerator\cite{stein2016robustness}. In the case of Ref. \onlinecite{stein2016robustness} a high magnetic field was also applied to improve the quantisation accuracy. These types of refrigerators carry significant cost challenges, and are not widely available at NMIs. Precision measurements of a pump at the temperature of liquid helium, or on multiple cool-downs, have not yet been reported.

In this work, we address some of these practical aspects of electron pumps. A well-characterised silicon electron pump was measured by three laboratories: NPL (UK), VTT MIKES (Finland) and Aalto University (Finland), designated throughout this work as `NPL', `MIKES' and `Aalto'. At each institute, the pump current was found to agree with the ideal error-free current $Ne \times f$, where $N=1$ is the number of electrons pumped in each cycle of the clock frequency $f$, within a relative uncertainty of $10^{-6}$ or less. Significantly, these precision measurements were performed with the pump cooled in liquid helium at a temperature of $4.2$~K, and zero applied magnetic field. These experimental conditions are considerably relaxed compared to all previous precision measurements. The measurements at MIKES and Aalto were also carried out using a quick and simple tuning procedure developed at NPL, yielding precision measurements $1-2$ days after cooling down the pump.

\section{Device and experiment time-line}

\begin{figure}
\includegraphics[width=9cm]{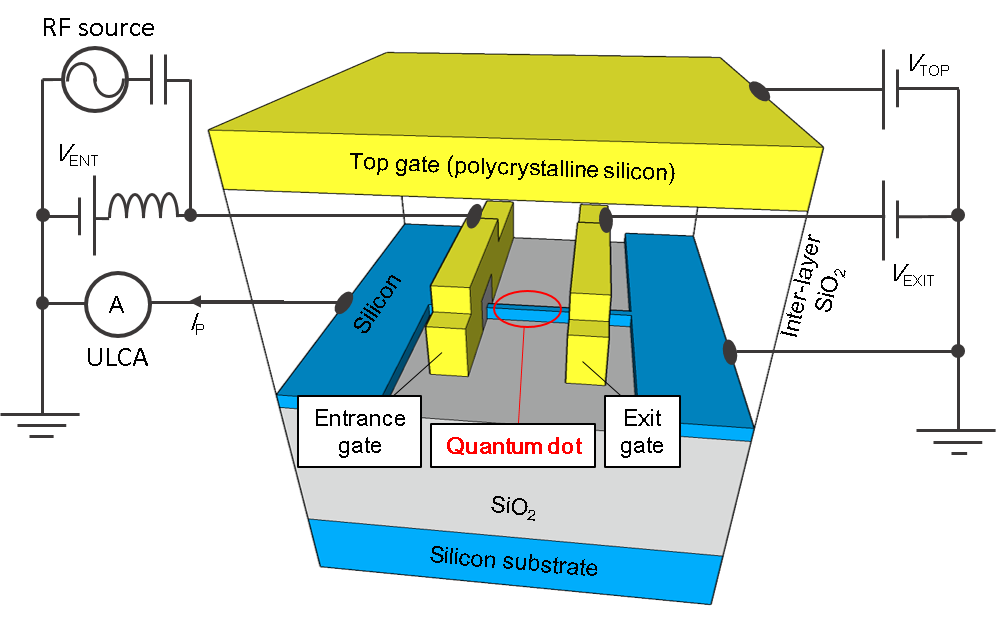}
\caption{\label{DeviceFig}\textsf{Schematic diagram of the electron pump device, showing the electrical connections to the device terminals. Electrons are pumped from left to right, so the ultrastable low-noise current amplifier (ULCA) connected to the left terminal measures current with a positive sign.}}
\end{figure}

The electron pump used in this work has been previously measured\cite{yamahata2016gigahertz} in 2015, when it demonstrated a pump current $I_{\text{P}}$ on the $N=1$ plateau equal to $ef$ within a relative uncertainty of $9.2 \times 10^{-7}$. The device, illustrated schematically in figure \ref{DeviceFig}, is a silicon nano-MOSFET, in which the charge carriers are induced in an undoped nano-wire by applying a positive voltage $V_{\text{TOP}} = 4$~V to a top gate\cite{fujiwara2008nanoampere,yamahata2016gigahertz}. Negative voltages $V_{\text{ENT}}$ and $V_{\text{EXIT}}$ applied to two finger gates crossing the nano-wire create potential barriers and define a quantum dot in between the barriers. Ratchet-mode single-electron pumping was induced by adding an AC signal to $V_{\text{ENT}}$ through a 3-dB attenuator and a room-temperature bias-tee. The AC pumping signal was a sine wave from an RF source with output power level denoted $P_{\text{RF}}$. 

The 2015 measurements\cite{yamahata2016gigahertz} were performed at a temperature of $1.5$~K, zero applied magnetic field, and $f=1$~GHz. In the intervening period, the device was left bonded into its NPL-designed sample holder, which was stored in an anti-static box at room temperature. In this work the device is cooled to a temperature of $4.2$~K by lowering it into a dewar of liquid helium. The same NPL-designed cryogenic probe is used to cool the sample at the three laboratories, but different liquid helium cryostats (also known as dewars) are used at each laboratory, and different models of commercial instrument are used to generate the DC and AC control voltages. The DC voltages $V_{\text{ENT}}$, $V_{\text{EXIT}}$ and $V_{\text{TOP}}$ were filtered using low-pass filters (not shown in figure \ref{DeviceFig}), with $2 \pi RC=14$~ms, to suppress noise from the electronic voltage sources. The same filters were used at all the laboratories. The DC wiring in the probe was of a custom design to minimise electrical noise due to vibration and triboelectric effects, and its design is summarised in supplementary section G. A pumping frequency of $f=1.05$~GHz is used for all the measurements, generating a current of $I_{\text{P}} \approx 168$~pA. The choice of frequency is constrained by dips in the transmission of the high-frequency wiring in the cryogenic probe.

The temporal order of the experiments is as follows: first, the stability of the pump under multiple cool-downs was evaluated at NPL. The pump was cooled down 7 times between July 2018 and February 2019, showing remarkable reproducibility. The values of $V_{\text{ENT}}$ and $V_{\text{EXIT}}$ for tuning the device to the one-electron plateau varied by less than $10 \%$ from one cool-down to the next. Precision measurements of the pump current were made on cool-downs 5 and 6, and data from both cool-downs is reported in this paper. During these cool-downs, the robustness of the pump current to changes in $V_{\text{ENT}}$ and $V_{\text{EXIT}}$ was evaluated, and a provisional tuning procedure was developed for rapidly locating the optimal values of these voltages for accurate pumping. The device was hand-carried by air to MIKES on 7th April 2019, and cooled down the next day. The rapid tuning procedure was applied, and the first precision measurements were made less than 24 hours after cooling the sample. Following a campaign of measurements at MIKES, the device was warmed up and hand-carried to Aalto (a distance of less than 1 km) on 5th May 2019. For this transportation, the device sample holder was not removed from the cryogenic probe. The tuning procedure was applied in the same way at Aalto, as at MIKES. The campaign of measurements at Aalto lasted until 3rd June.

\section{precision measurement setups}

All measurements of the pump current are carried out using an ultrastable low-noise current amplifier (ULCA)\cite{drung2015ultrastable} connected to the source side of the pump (see figure \ref{DeviceFig}). This is a transresistance amplifier with nominal current-to-voltage gain $A_{\text{TR}} = 10^9$~V/A, with the key feature that this gain is very stable in time: the gain of a number of ULCA units has been shown to be stable at the level of 1 part in $10^{6}$ on time-scales of a year\cite{krause2019noise}. A single ULCA unit (the `NPL ULCA') is used for measurements at NPL. Its gain is calibrated using the NPL high-resistance CCC\cite{giblin2019interlaboratory}. A second ULCA unit (the `MIKES ULCA') is used for measurements at MIKES and Aalto. Its gain is calibrated at MIKES using a Magnicon CCC. The ULCA was transported between MIKES and Aalto by car. Although the temperature of the ULCA was not logged during these short transportations, they took place during a warm time of year. Shifts in the ULCA gain which have been observed when the ULCA is exposed to low temperatures during transportation \cite{giblin2019interlaboratory,krause2019noise} are not expected to be a problem here. More detailed information about the ULCA calibrations can be found in the supplementary information, section B. 

The output voltage of the ULCA is digitised by a precision digital voltmeter (DVM). All three laboratories used a Hewlett Packard / Agilent / Keysight 3458A for this function\footnote{Mention of specific models of commercial instrument is for information only and does not imply endorsement by the authors or their respective institutions}. The DVM is set to integrate each data point for 10 power line cycles (PLC), with an auto zero operation every 20 data points. The Optimisation of the DVM auto zero is discussed in Ref. \onlinecite{stein2016robustness}. Each laboratory uses a different DVM unit. All DVMs are calibrated with traceability to a Josephson voltage standard (JVS), but the exact traceability routes are different. At NPL, the DVM is calibrated directly against a JVS using an automated switch with a calibration interval of approximately an hour. At MIKES, some calibrations are carried out directly against the JVS, and some using an intermediate 1-V Zener voltage standard. The minimum calibration interval is approximately a day. At Aalto, all calibrations are carried out using a Zener voltage standard calibrated against the JVS at MIKES and hand-carried to Aalto. This resulted in a higher uncertainty contribution due to the DVM calibration at Aalto, shown in table \ref{UncertTable}.

In this paper we present two types of data: for `standard precision' data, the ULCA output voltage is recorded as a pump parameter such as a gate voltage is scanned. This type of data is used for characterisation of the pump. For `high precision' data, as in previous studies, the pump is turned on and off, and $I_{\text{P}}$ is extracted from the on-off difference signal to eliminate possible instrumental offset drift from the data. Two types of on-off cycle are used in this work. In the `power switching' cycle, the AC drive signal at the entrance gate is turned on and off. This is how all previous precision pump measurements have been carried out\cite{giblin2019evidence}, although here we use a longer on-off cycle than typically used in the past: 1000 data points (228 seconds) per on or off segment at NPL and MIKES, 1300 points (296 seconds) per segment at Aalto. The first 300 points were rejected from each data segment prior to analysis, to avoid time constant effects. Note that the time given for each segment includes the time required for the auto zero operations. In the `gate switching' cycle, the AC drive is left on, and $V_{\text{EXIT}}$ is stepped from its operation point to $-1.7$~V, well into the $N=0$ region of the pump map. This type of cycle avoids time constants possibly due to RF heating (see supplementary section D) but $I_{\text{P}}$ has to be corrected for the change in leakage current due to stepping the gate voltage. This correction is also detailed in the supplementary information, section E. As shown in table I, the uncertainty in the leakage correction is the largest contribution to the combined uncertainty in the measurements using the gate switching cycle.

A note on the data analysis: Our main results are reported as dimensionless numbers, the fractional deviation of the pump current from $ef$: $\Delta I_{\text{P}} = (I_{\text{P}} - ef )/ (ef)$, and are therefore independent of the choice of unit system. Since the experiments pre-date the May 2019 redefinition of the SI, we chose to analyse the precision measurements within the system of 1990 electrical units: The calibration of the ULCA gain is traced to the quantum Hall resistance using $R_{\text{K-90}}$, the voltmeters are calibrated with reference to $K_{\text{J-90}}$, and we report the deviation of the pump current from $f \times e_{\text{90}}$, where $e_{\text{90}} \equiv 2/(K_{\text{J-90}} R_{\text{K-90}})$. The constants $R_{\text{K-90}}$ and $K_{\text{J-90}}$ are the fixed values assigned to the von Klitzing and Josephson constants respectively in 1990. They were used for representing the SI ohm and volt from then until the 2019 redefinition. 

\begin{figure}
\includegraphics[width=9cm]{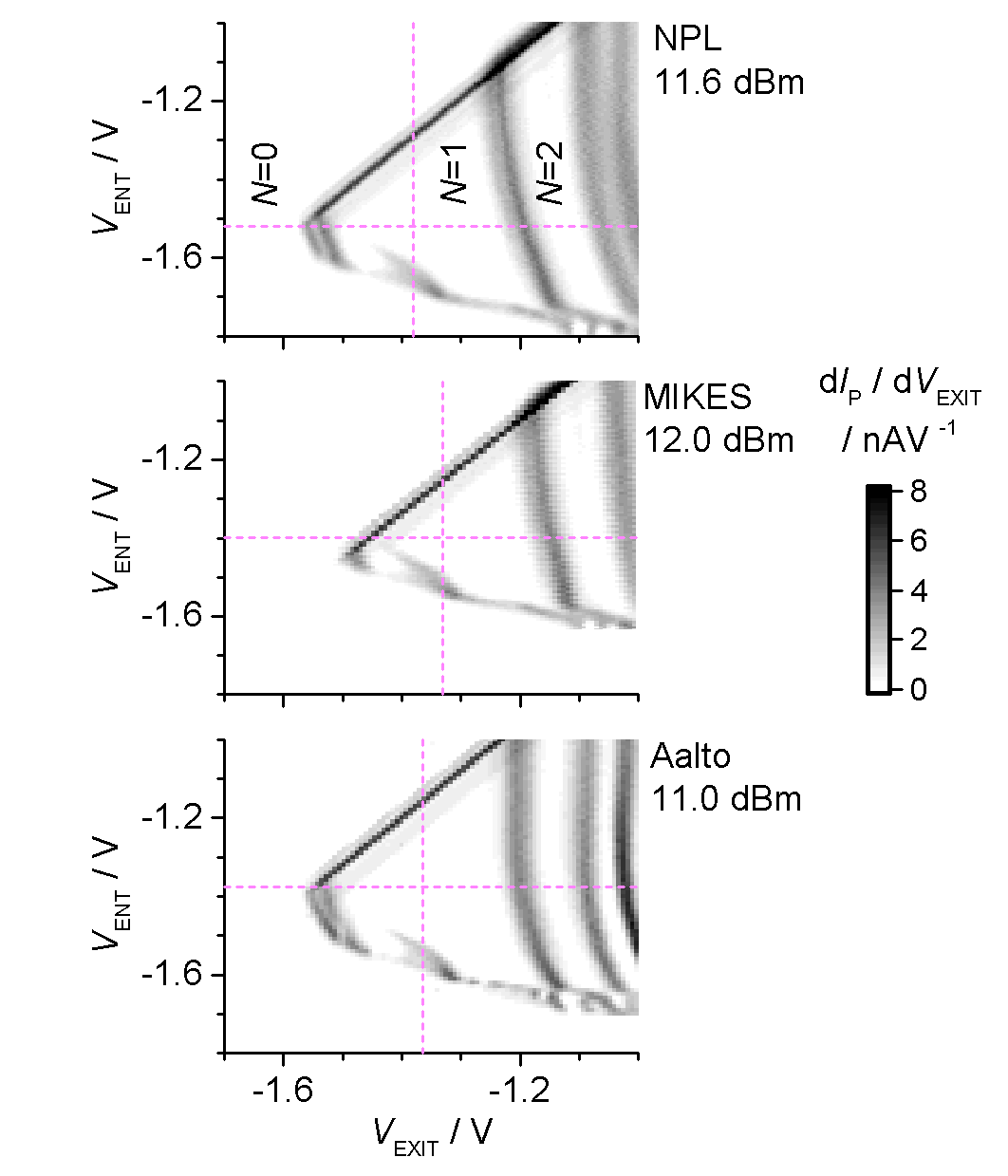}
\caption{\label{PumpMapFig}\textsf{Derivative of the pump current with respect to exit gate voltage, as a function of entrance and exit gate voltages, measured at NPL, MIKES, and Aalto. The number of electrons pumped per cycle, $N$, is indicated in the first three plateau regions of the top-most panel. The laboratory identifier and RF generator output power $P_{\text{RF}}$ are indicated next to each panel. The data of figure \ref{LogScalePlots} was taken along the cuts indicated by the horizontal and vertical dashed lines.}}
\end{figure}

\section{characterisation and tuning}

After cooling down the device in liquid helium, the first stage of characterisation is to record a pump map. In figure \ref{PumpMapFig}, we present differential pump current maps measured at the three laboratories. The pump map is a standard fingerprint for a tunable-barrier pump, which shows the ranges of $V_{\text{ENT}}$ and $V_{\text{EXIT}}$ where the current is quantised (white areas) and the transitions between quantised plateaus (black lines). The characteristic pattern of quantised current regions establishes that the pump is operating in a ratchet mode\cite{kaestner2008robust}. The similarity between the three pump maps shows the stability of the pump after thermal cycling and transportation, but some small differences are visible. Despite a higher RF generator power, the pump map measured at MIKES is not as extended along the $V_{\text{ENT}}$ axis as the map measured at NPL, indicating a lower AC voltage present on the entrance gate. Similar behaviour has also been observed during multiple cool-downs at NPL. This variation may be due to two processes: firstly, small changes in carrier concentration of the sample and secondly, changes in the reflection co-efficients at connectors in the coaxial RF transmission line. The latter is a plausible mechanism, as the properties of the RF connectors are sensitive to mechanical strain induced by the large temperature gradient along the transmission line. The pump occasionally switched to a different state characterised by a wider pump map along the entrance gate axis, as detailed in supplementary section F. None of the measurements reported in the paper were carried out in this state.

\begin{figure}
\includegraphics[width=9cm]{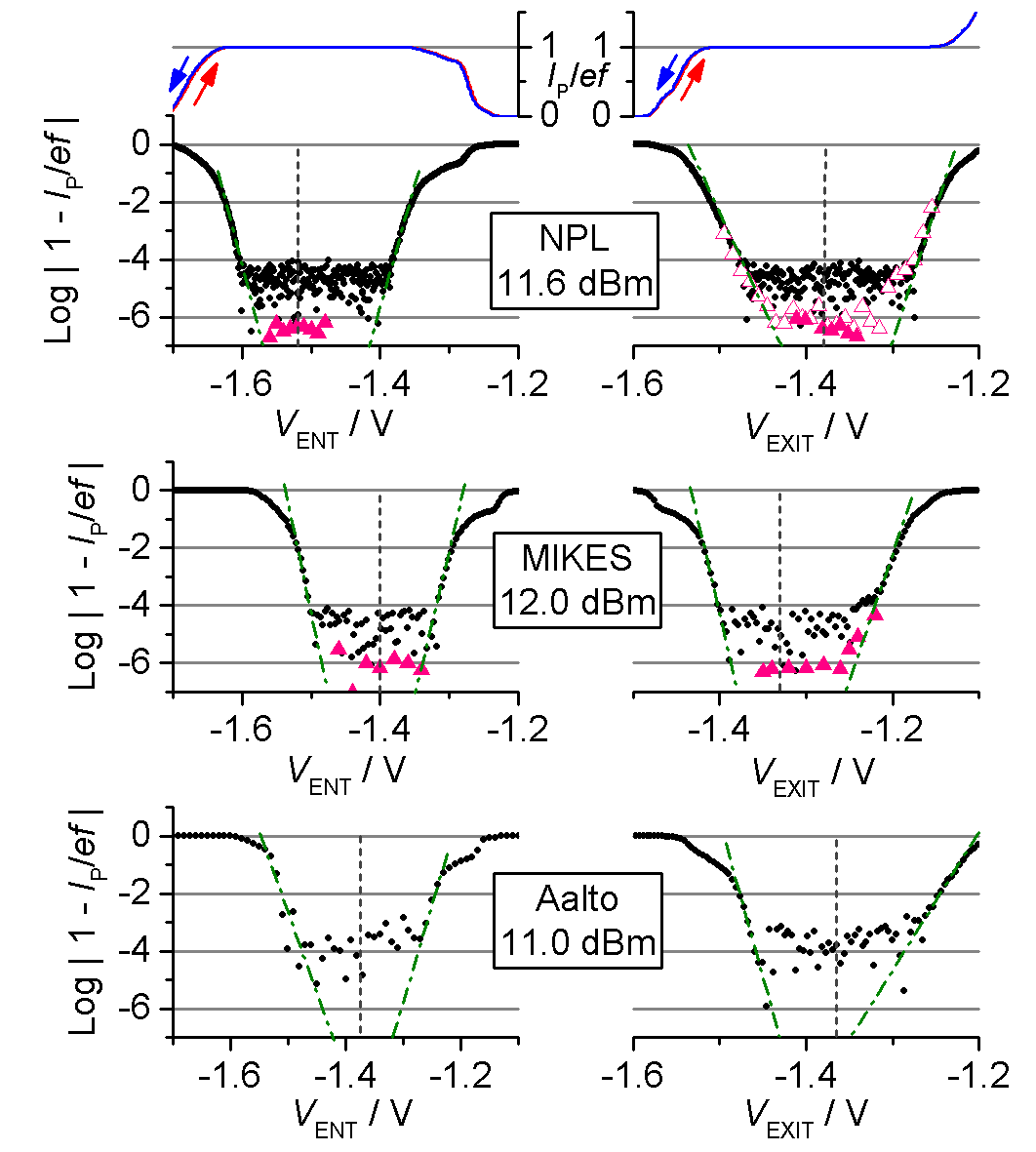}
\caption{\label{LogScalePlots}\textsf{The 6 main plots show relative deviation of the pump current from the expected value $e \times f$ along the $V_{\text{ENT}}$ and $V_{\text{EXIT}}$ axes indicated by the dashed lines in figure \ref{PumpMapFig}. Black circles: Standard precision data, where each data point is a single $10$~PLC measurement. Data was taken by sweeping the voltages from negative to positive (left to right along the x-axis). The dash-dot lines are guides to the eye for the exponential approaches to the $N=1$ plateau. Open and closed pink triangles: High-precision data, where each data point is extracted from a number of on-off cycles. The laboratory identifier and RF generator power $P_{\text{RF}}$ are indicated next to each pair of plots.  Vertical dashed lines indicate the values of $V_{\text{ENT}}$ and $V_{\text{EXIT}}$ selected for the high-precision measurements presented in figure \ref{SummaryHighAccFig}. The two small plots at the top of the figure show the NPL standard-precision data on linear y-axes (red lines), together with additional data (blue lines) taken with the sweep direction reversed.}}
\end{figure}

Next, the extent of the quantised current region is estimated by studying the deviation of $I_{\text{P}}$ from $ef$ on a logarithmic scale\cite{giblin2017robust}, in figure \ref{LogScalePlots}, along the line cuts indicated by dotted lines on the pump maps of figure \ref{PumpMapFig}. The fixed values of $V_{\text{ENT}}$ and $V_{\text{EXIT}}$ for these line cuts are established from a few iterations of plotting data similar to figure \ref{LogScalePlots}, starting from initial guesses, and adjusting $V_{\text{ENT}}$ and $V_{\text{EXIT}}$ with each iteration to maximise the plateau width. The exponential approaches to the plateau can be extrapolated (dash-dot lines) to predict the extent of the plateau at the $0.1$~ppm accuracy level. At NPL, high-precision scans (open and closed pink triangles) were carried out to verify the flatness of the plateaus at the 1-ppm level, before selecting an operation point, $V_{\text{ENT}} = -1.52$~V and $V_{\text{EXIT}} = -1.38$~V, indicated by the vertical dashed lines. At MIKES and Aalto, the emphasis is on a quick characterisation procedure, and operation points are chosen based solely on standard-precision data. A certain amount of subjective judgment entered into the selection of these operation points. For example, the exit gate scan at MIKES seems to indicate a shoulder, only just resolved by the black data points, at $V_{\text{EXIT}}\sim -1.25$~V. This feature biases the selection of the operation point towards more negative $V_{\text{EXIT}}$. The feature does not appear in the high-precision scan (pink triangles), but this scan was measured after the long high-precision measurements shown in figure \ref{SummaryHighAccFig}, and did not play a role in the selection of the operation point. We also checked that the location of the plateau was not biased by the finite scan rate of the gate voltage combined with possible hysteresis, by comparing scans with opposite scan directions. Two such pairs of scans are shown above the NPL log-scale plots, with the same x-axis scales and linear y-axes, showing negligible hysteresis.

\section{high-precision measurements}

\begin{figure}
\includegraphics[width=9cm]{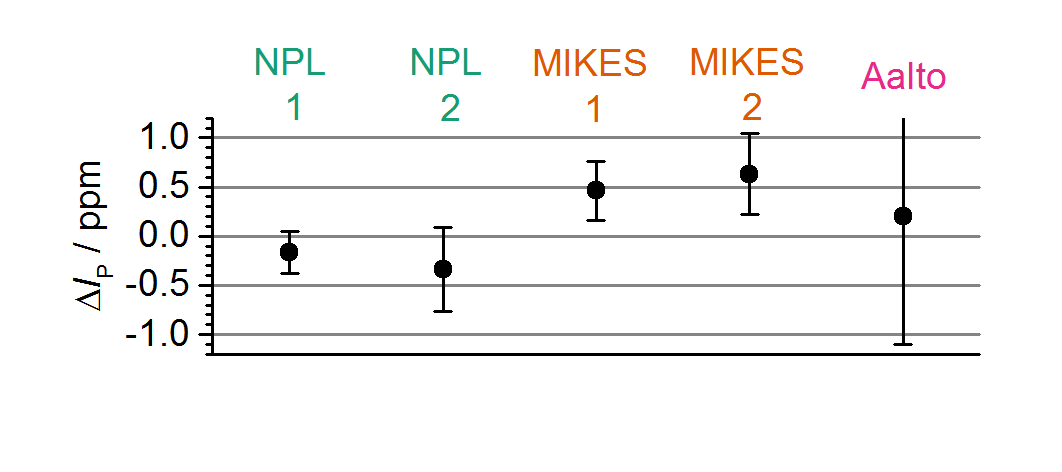}
\caption{\label{SummaryHighAccFig}\textsf{Results of long high-precision measurements at fixed gate voltages, expressed as the dimensionless deviation of the pump current from $e_{\text{90}}f$. Error bars are combined standard uncertainties ($k=1$). Entrance and exit gate voltages for the data points `NPL 1', `MIKES 1' and `Aalto' are shown by the intersections of the dashed lines in figure \ref{PumpMapFig}. data points `NPL 2' and `MIKES 2' were taken at slightly different gate voltages following a repeat of the tuning procedure.}}
\end{figure}

At each laboratory, several high-precision measurements are carried out at the optimal operation points determined from figure \ref{LogScalePlots}. These measurements are typically carried out overnight, with averaging times from $10$ to $22$ hours. Results from five of them are shown in figure \ref{SummaryHighAccFig}, in terms of the dimensionless deviation of the pump current from $e_{\text{90}}  f$, $\Delta I_{\text{P}} = (I_{\text{P}} - e_{\text{90}}f )/ (e_{\text{90}}f)$. Measurements denoted NPL 1, MIKES 1, and Aalto utilise the power switching on-off cycle, and measurements NPL 2 and MIKES 2 utilise the gate switching cycle. Error bars indicate the combined standard (k=1) uncertainty.

\begin{table*}
\caption{\label{UncertTable} Breakdown of the uncertainty components for the five long high-precision measurements shown in figure \ref{SummaryHighAccFig}. All entries in the table are dimensionless relative uncertainties ($k=1$) in parts per million. The largest uncertainty contribution for each measurement is highlighted in bold type.}
\centering
\setlength{\tabcolsep}{8pt}
\begin{tabular}{c c c c c c}
Contribution & NPL 1 & NPL 2 & MIKES 1 & MIKES 2 & Aalto \\
\hline\hline
ULCA $G_{\text{I}}$ Calibration & 0.023 & 0.031 & 0.035 & 0.044 & 0.044 \\ 
ULCA $R_{\text{IV}}$ Calibration & 0.056 & 0.056 & 0.065 & 0.066 & 0.066 \\ 
ULCA $G_{\text{I}}$ Drift & 0.014 & 0.014 & 0.044 & 0.074 & 0.074 \\ 
ULCA $R_{\text{IV}}$ Drift & 0.019 & 0.019 & 0.018 & 0.042 & 0.042 \\
Voltmeter Calibration & - & - & 0.17 & 0.18 & \textbf{1.20} \\
Voltmeter Drift & - & - & \textbf{0.18} & 0.046 & 0.38 \\
Leakage correction & - & \textbf{0.40} & - & \textbf{0.30} & - \\
Pump measurement type A  & \textbf{0.20} & 0.14 & 0.137 & 0.17 & 0.25 \\
\hline
Total & 0.21 & 0.43 & 0.30 & 0.41 & 1.30 \\
\end{tabular}
\end{table*}

\begin{figure}
\includegraphics[width=9cm]{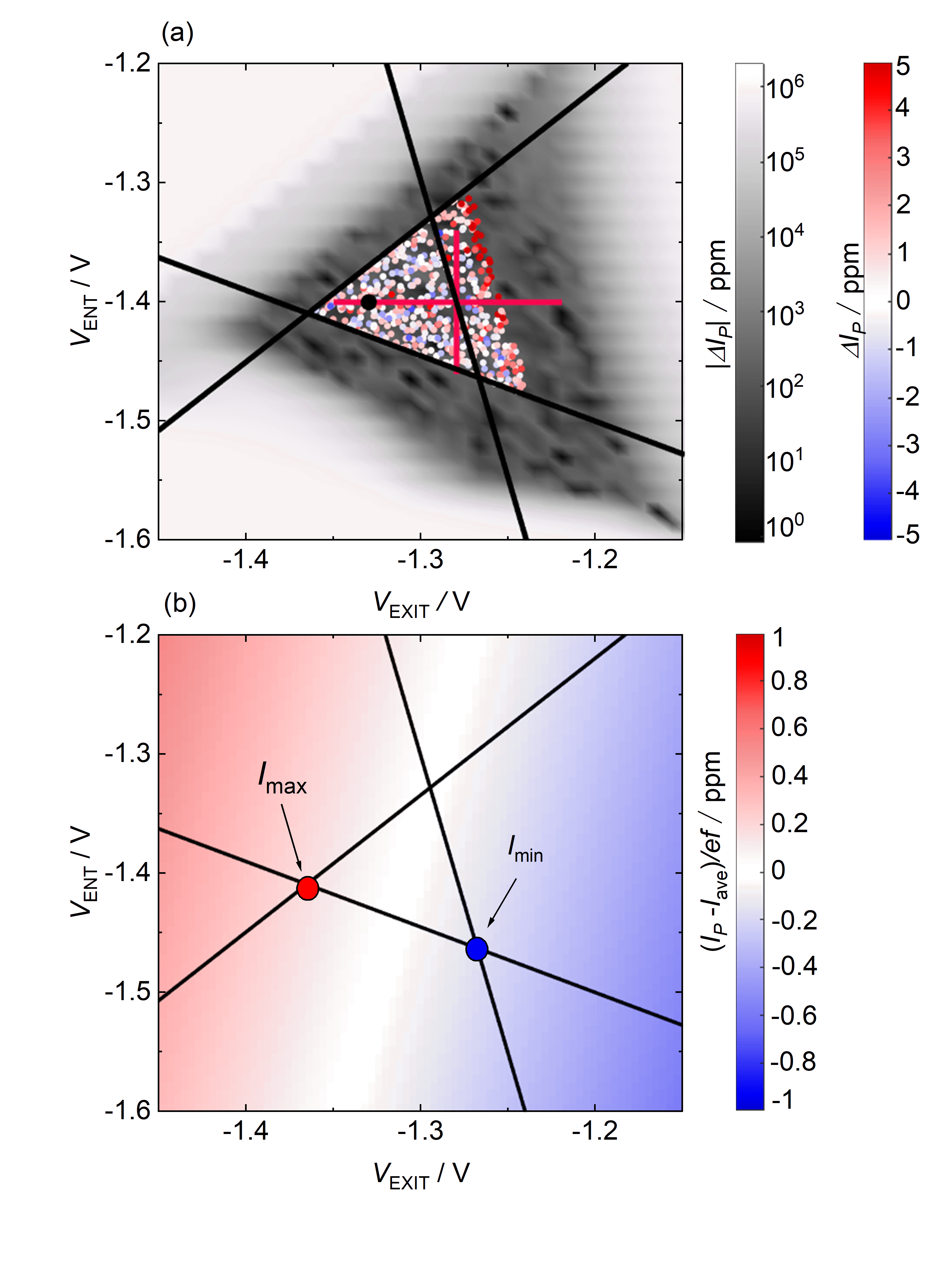}
\caption{\label{random_sampling}\textsf{Method for studying the robustness of single-electron pumps. The triangular data range used for the multiple-linear-regression fit is shown by the black lines. (a) Standard precision data of the quantized current plateau is shown on greyscale. Randomized high-precision current measurement data are shown by coloured dots. The magenta lines show the parameter regime where the traceable MIKES data of Fig.~\ref{LogScalePlots} are measured. The black dot shows where the precision measurement of Fig.~\ref{SummaryHighAccFig} was carried out. (b) The results of the multiple-linear-regression analysis of the current plateau inside the triangular area between the black lines. The deviation of the fitted plane from average value $I_\text{ave}=(I_\text{max}+I_\text{min})/2$ is plotted using a colour code to illustrate the small tilt of the fit plane. $I_\text{max}$ and $I_\text{min}$ are the maximum and minimum values of the fitted currents inside the fit area, respectively.}}
\end{figure}

The breakdown of the uncertainties is detailed in table \ref{UncertTable}. We have distinguished the uncertainties due to the two separate stages of calibrating the ULCA: The input current gain stage $G_{\text{I}}$, and the output trans-resistance stage $R_{\text{IV}}$. At NPL, the DVM calibrations are interleaved with the pump measurements, so that the uncertainty in these calibrations does not appear as separate terms: instead it is combined into the type A uncertainty as detailed in supplementary section C. At MIKES and Aalto, the DVM calibrations performed before and after the pump measurement give separate uncertainty contributions due to DVM calibration uncertainty and drift. The leakage correction, described in supplementary section E, only contributes an uncertainty to measurements using the gate-switching on-off cycle. Terms contributing a relative uncertainty less than $10^{-8}$, for example the uncertainty in the pump frequency $f$, have been neglected from the analysis.

\section{plateau robustness}

In recent years, there has been discussion\cite{giblin2019evidence,scherer2019single} about the possible form of guidelines for testing the accuracy of single electron pumps, in analogy to the guidelines for quantum Hall resistance standards\cite{delahaye2003revised}. A key conclusion is that a candidate single-electron-based quantum current standard should demonstrate robustness of the pump current. This means that the current is independent of all control parameters within a parameter space that is experimentally feasible. The control parameters include (but are not limited to) DC gate voltages, and RF power. Experimental feasibility means that the control parameters do not have to be tuned with an unreasonably high level of precision, and that fluctuations in the tuning parameters, for example due to the imperfections of electronic sources, do not significantly change the output current.

The typical method for testing robustness, used in several studies thus far\cite{stein2016robustness,giblin2017robust,zhao2017thermal}, reviewed in Ref. \onlinecite{giblin2019evidence} and also shown in Fig. \ref{LogScalePlots}, is to find an optimal operation point and then vary the control parameters one at a time while carrying out traceable precision measurements. This method has two main limitations. Firstly, it measures the robustness along lines in a multidimensional parameter space, which samples only a small fraction of the total parameter space. Secondly, it is also typical practice to collect the data in sequential order, incrementing the stepped control parameter from its minimum value to its maximum value. Sequential stepping has the problem that any time correlation of the measurement data (for example, caused by a drift in the calibration factor of the ULCA or DVM) will yield a spurious parameter space correlation of the data. Here, we investigated a technique which avoids these two limitations.

In figure~\ref{random_sampling}, we present colour-map data of the plateau flatness in the two-dimensional parameter space defined by $V_{\text{ENT}}$ and $V_{\text{EXIT}}$. Each coloured dot is the result of a single on-off cycle using the power switching method described in section III. The values of $V_{\text{ENT}}$ and $V_{\text{EXIT}}$ for each data point are selected randomly, with the constraint that they should lie within a triangular region which, on a coarse scale, defines the plateau. Our measurement scheme can be generalized to all control parameters, but these two enable us to demonstrate the method.

The measurement of Fig.~\ref{random_sampling} took 149 hours. Neither the voltmeter nor the ULCA were calibrated during this time, so the measurement is not traceable. However, in this experiment we are only concerned with the flatness of the plateau, not its absolute value. From past experience, we expect the gain of the voltmeter to drift by up to 1 ppm on the time scale of the measurement, but we cannot make any assumptions about the magnitude or frequency spectrum of the drift. The key point is that any changes in the gain of the measurement system, whether correlated in time or not, will appear as uncorrelated fluctuations along the $V_{\text{ENT}}$ and $V_{\text{EXIT}}$ axes and will be indistinguishable from fluctuations due to random noise. The effect of the changes in measurement system gain will be to increase the uncertainty in the fitted plateau slope, but the value of the fitted slope itself will only contain information about dependence of the pump current on $V_{\text{ENT}}$ and $V_{\text{EXIT}}$. 

Linear regression analysis (fitting a plane to a 2-dimensional data set) was performed on a subset of the precision data of Fig.~\ref{random_sampling}a, bounded by the thick black lines in the figure. The data subset was chosen to avoid the clear increase in the pump current at the right of the plot. The standard deviation of the data points used for the fit is 1.5~ppm. The fitted control parameter dependencies of $I_{\text{P}}$ are $(-3.0\pm 4.3)$~ppm/V and $(0.5\pm 3.2)$~ppm/V for $V_{\text{EXIT}}$ and $V_{\text{ENT}}$, respectively. The plateau tilt $\Delta I =I_\mathrm{max}-I_\mathrm{min}=(0.3\pm 0.5)$~ppm. This numerical statement about the plateau flatness is essentially a 2-dimensional extension of earlier attempts to define the pump plateau as the range of data points in a one-dimensional scan for which the uncertainty in the gradient of a linear fit is greater than the fitted gradient\cite{giblin2017robust}. In this work, unlike in Ref. \onlinecite{giblin2017robust}, we have randomised the order of the data points which means that drift in the calibration of the measuring instruments only increases the random scatter of the data points (and therefore the uncertainty in the fitted slope), but cannot be mistaken for a sloped plateau. We find that there is thus no tilt of the plateau within the measurement uncertainty. Finally, we want to point out that in a more optimized measurement protocol, which does not include zero measurements or too many points outside the accurate plateau, a similar uncertainty for the robustness can be obtained in a few days.

\section{discussion}

The electron pump used in this study displayed, overall, a remarkable level of stability over many cooldowns and handling procedures. The stability of Coulomb blockade features of silicon single-electron devices is already well-known \cite{zimmerman2001excellent,hourdakis2008lack}, and we discovered that this stability carries onto high-precision electron pumping. The measurements at MIKES and Aalto showed that a rapid characterisation procedure could yield a pump current accurate to a part per million, on a measurement time-scale of roughly 2 days. This is comparable to the time required for basic checks on a quantum Hall resistance sample\cite{delahaye2003revised} prior to using it as a primary resistance standard. The tuning procedure detailed in section IV was empirically developed based on approximately 3 weeks of measurement data taken at NPL. We are not proposing it as a universal method for tuning an electron pump, but our data provide encouraging evidence that such a universal and useful method may be developed.  

The accurate operation of the pump at a temperature of $4$~K is one of the key findings of this study. This is due to the large addition energy of the silicon nanowire quantum dot. We did not estimate the addition energy for our particular device, but recent measurements of a similar device resulted in an estimate $E_{\text{add}} = 12$~meV\cite{yamahata2019picosecond}, corresponding to a characteristic temperature $E_{\text{add}} / k_{\text{B}} = 140$~K. The large addition energy for the silicon nanowire pumps is also shown by the presence of current plateaus, in a similar device to the one used in this study, at a temperature of $20$~K and $f=2.3$~GHz \cite{fujiwara2008nanoampere}. In contrast, typical GaAs quantum dots have charging energies in the range $1$ to $2$~meV\cite{gerster2019robust}, and the most accurate measurements on these pumps have been done in a dilution refrigerator at temperatures of $0.1$~K\cite{stein2015validation,stein2016robustness}. In previous studies, fits to the $I_{\text{P}} (V_{\text{EXIT}})$ data have determined whether back-tunneling\cite{giblin2017robust}, or thermal exchange of electrons with the source lead\cite{zhao2017thermal} was the dominant mechanism in determining the number of electrons pumped in each cycle. It was not possible to extract this information for the device in this study, because of the anomalous transition from pumping zero to one electrons, clearly visible as the double line at the `nose' of the NPL derivative pump map in figure \ref{PumpMapFig}. Further studies will investigate the pumping mechanism, as well as probing the upper frequency limit for accurate pumping.

The MIKES results on the pump current are above $ef$ by marginally more than a standard deviation. This offset is on the border of statistical significance, but it is noteworthy that it appears in both the MIKES 1 and MIKES 2 runs, which employed different types of on-off cycling, and in other precision measurements at different operation points (not shown) made during this cool-down at MIKES. Thus the offset is unlikely to be due to an error in the leakage current correction, which would affect MIKES 2 only. It is also unlikely to be due to an error in the voltmeter calibration, as the measurement runs were spanned by several voltmeter calibrations, some directly against a Josephson array and some using a Zener standard. The cause of the offset is under investigation. One possibility is an error in the gain of the MIKES ULCA. The calibration history of the input current gain $G_{\text{I}}$ of the MIKES ULCA shows steplike changes with relative magnitude larger than $0.2$ ppm (see supplementary section B) Such changes in between the ULCA calibration and the pump current measurements would directly affect the measurement results. If this was the case, the Aalto measurements, which used the same ULCA unit, would also be offset, but the larger uncertainty in the Aalto measurements due to the DVM calibration means that the offset cannot be resolved. Another possibility is an error in the calibration of the ULCA output stage gain $R_{\text{IV}}$ which has a nominal value of $1$~M$\Omega$. It is unlikely that a $0.5$~ppm error would appear in CCC-based traceabilily to $1$~M$\Omega$, but a bilateral comparison of stable standard resistors such as the one reported in Ref. \onlinecite{giblin2019interlaboratory} could resolve this question.

All the presented results were obtained on a single device, which was extensively characterised and the subject of earlier precision measurements \cite{yamahata2016gigahertz}. At the time this study was conceived, this device was the most promising one available for operation at liquid helium temperature, although some other devices showed similar pumping characteristics in characterisation measurements. Improving the yield of fabrication processes such that devices with good performance become readily available is clearly an important problem which requires further work. Some progress has already been made: for GaAs quantum dots, a recent systematic study has established clear correlations between fabrication geometry and quantum dot properties such as the charging energy \cite{gerster2019robust}. This study did not investigate pumping, only static quantum dot properties accessible from DC measurements. The authors are not aware of any analogous study for silicon devices. Most groups working in the field report anecdotally that there is considerable variation in pumping performance from a batch of devices with nominally identical fabrication properties, although no systematic study has yet been published on this subject. It will be an important direction for future work.

\section{conclusions}

We have reported measurements of the same electron pump at three different institutes, with uncertainties of roughly a part per million or less. All three sets of measurements are broadly in agreement with the ideal error-free current $ef$ within the uncertainty. These are the first traceable high-precision measurements of an electron pump at the relatively high temperature of liquid helium, and they demonstrate that a well-characterised electron pump can operate as low-current reference standard using resources (a liquid helium dewar, RF sine wave synthesiser, DC voltage sources and a frequency reference) commonly available in National Measurement Institutes.

\begin{acknowledgments}
This research was supported by the UK department for Business, Energy and Industrial Strategy and the EMPIR Joint Research Project `e-SI-Amp' (15SIB08). The European Metrology Programme for Innovation and Research (EMPIR) is co-financed by the Participating States and from the European Union's Horizon 2020 research and innovation programme. M. M. and M. J. acknowledge the Academy of Finland for funding through its Centre of Excellence in Quantum Technology (QTF), project no. 312300. A.F. and G.Y. are supported by JSPS KAKENHI Grant Number JP18H05258.
\end{acknowledgments}

\bibliography{SPGrefs_ComparisonPaper}

\begin{thebibliography}{23}%
\makeatletter
\providecommand \@ifxundefined [1]{%
 \@ifx{#1\undefined}
}%
\providecommand \@ifnum [1]{%
 \ifnum #1\expandafter \@firstoftwo
 \else \expandafter \@secondoftwo
 \fi
}%
\providecommand \@ifx [1]{%
 \ifx #1\expandafter \@firstoftwo
 \else \expandafter \@secondoftwo
 \fi
}%
\providecommand \natexlab [1]{#1}%
\providecommand \enquote  [1]{``#1''}%
\providecommand \bibnamefont  [1]{#1}%
\providecommand \bibfnamefont [1]{#1}%
\providecommand \citenamefont [1]{#1}%
\providecommand \href@noop [0]{\@secondoftwo}%
\providecommand \href [0]{\begingroup \@sanitize@url \@href}%
\providecommand \@href[1]{\@@startlink{#1}\@@href}%
\providecommand \@@href[1]{\endgroup#1\@@endlink}%
\providecommand \@sanitize@url [0]{\catcode `\\12\catcode `\$12\catcode
  `\&12\catcode `\#12\catcode `\^12\catcode `\_12\catcode `\%12\relax}%
\providecommand \@@startlink[1]{}%
\providecommand \@@endlink[0]{}%
\providecommand \url  [0]{\begingroup\@sanitize@url \@url }%
\providecommand \@url [1]{\endgroup\@href {#1}{\urlprefix }}%
\providecommand \urlprefix  [0]{URL }%
\providecommand \Eprint [0]{\href }%
\providecommand \doibase [0]{http://dx.doi.org/}%
\providecommand \selectlanguage [0]{\@gobble}%
\providecommand \bibinfo  [0]{\@secondoftwo}%
\providecommand \bibfield  [0]{\@secondoftwo}%
\providecommand \translation [1]{[#1]}%
\providecommand \BibitemOpen [0]{}%
\providecommand \bibitemStop [0]{}%
\providecommand \bibitemNoStop [0]{.\EOS\space}%
\providecommand \EOS [0]{\spacefactor3000\relax}%
\providecommand \BibitemShut  [1]{\csname bibitem#1\endcsname}%
\let\auto@bib@innerbib\@empty
\bibitem [{\citenamefont {Kaneko}, \citenamefont {Nakamura},\ and\
  \citenamefont {Okazaki}(2016)}]{kaneko2016review}%
  \BibitemOpen
  \bibfield  {author} {\bibinfo {author} {\bibfnamefont {N.-H.}\ \bibnamefont
  {Kaneko}}, \bibinfo {author} {\bibfnamefont {S.}~\bibnamefont {Nakamura}}, \
  and\ \bibinfo {author} {\bibfnamefont {Y.}~\bibnamefont {Okazaki}},\
  }\bibfield  {title} {\enquote {\bibinfo {title} {A review of the quantum
  current standard},}\ }\href@noop {} {\bibfield  {journal} {\bibinfo
  {journal} {Measurement Science and Technology}\ }\textbf {\bibinfo {volume}
  {27}},\ \bibinfo {pages} {032001} (\bibinfo {year} {2016})}\BibitemShut
  {NoStop}%
\bibitem [{\citenamefont {Scherer}\ and\ \citenamefont
  {Schumacher}(2019)}]{scherer2019single}%
  \BibitemOpen
  \bibfield  {author} {\bibinfo {author} {\bibfnamefont {H.}~\bibnamefont
  {Scherer}}\ and\ \bibinfo {author} {\bibfnamefont {H.~W.}\ \bibnamefont
  {Schumacher}},\ }\bibfield  {title} {\enquote {\bibinfo {title}
  {{Single-electron pumps and quantum current metrology in the revised SI}},}\
  }\href@noop {} {\bibfield  {journal} {\bibinfo  {journal} {Annalen der
  Physik}\ ,\ \bibinfo {pages} {1800371}} (\bibinfo {year} {2019})}\BibitemShut
  {NoStop}%
\bibitem [{\citenamefont {Pekola}\ \emph {et~al.}(2013)\citenamefont {Pekola},
  \citenamefont {Saira}, \citenamefont {Maisi}, \citenamefont {Kemppinen},
  \citenamefont {M{\"o}tt{\"o}nen}, \citenamefont {Pashkin},\ and\
  \citenamefont {Averin}}]{pekola2013single}%
  \BibitemOpen
  \bibfield  {author} {\bibinfo {author} {\bibfnamefont {J.~P.}\ \bibnamefont
  {Pekola}}, \bibinfo {author} {\bibfnamefont {O.-P.}\ \bibnamefont {Saira}},
  \bibinfo {author} {\bibfnamefont {V.~F.}\ \bibnamefont {Maisi}}, \bibinfo
  {author} {\bibfnamefont {A.}~\bibnamefont {Kemppinen}}, \bibinfo {author}
  {\bibfnamefont {M.}~\bibnamefont {M{\"o}tt{\"o}nen}}, \bibinfo {author}
  {\bibfnamefont {Y.~A.}\ \bibnamefont {Pashkin}}, \ and\ \bibinfo {author}
  {\bibfnamefont {D.~V.}\ \bibnamefont {Averin}},\ }\bibfield  {title}
  {\enquote {\bibinfo {title} {Single-electron current sources: Toward a
  refined definition of the ampere},}\ }\href@noop {} {\bibfield  {journal}
  {\bibinfo  {journal} {Reviews of Modern Physics}\ }\textbf {\bibinfo {volume}
  {85}},\ \bibinfo {pages} {1421} (\bibinfo {year} {2013})}\BibitemShut
  {NoStop}%
\bibitem [{\citenamefont {Kaestner}\ and\ \citenamefont
  {Kashcheyevs}(2015)}]{kaestner2015non}%
  \BibitemOpen
  \bibfield  {author} {\bibinfo {author} {\bibfnamefont {B.}~\bibnamefont
  {Kaestner}}\ and\ \bibinfo {author} {\bibfnamefont {V.}~\bibnamefont
  {Kashcheyevs}},\ }\bibfield  {title} {\enquote {\bibinfo {title}
  {Non-adiabatic quantized charge pumping with tunable-barrier quantum dots: a
  review of current progress},}\ }\href@noop {} {\bibfield  {journal} {\bibinfo
   {journal} {Reports on Progress in Physics}\ }\textbf {\bibinfo {volume}
  {78}},\ \bibinfo {pages} {103901} (\bibinfo {year} {2015})}\BibitemShut
  {NoStop}%
\bibitem [{\citenamefont {Giblin}\ \emph
  {et~al.}(2019{\natexlab{a}})\citenamefont {Giblin}, \citenamefont {Fujiwara},
  \citenamefont {Yamahata}, \citenamefont {Bae}, \citenamefont {Kim},
  \citenamefont {Rossi}, \citenamefont {M{\"o}tt{\"o}nen},\ and\ \citenamefont
  {Kataoka}}]{giblin2019evidence}%
  \BibitemOpen
  \bibfield  {author} {\bibinfo {author} {\bibfnamefont {S.}~\bibnamefont
  {Giblin}}, \bibinfo {author} {\bibfnamefont {A.}~\bibnamefont {Fujiwara}},
  \bibinfo {author} {\bibfnamefont {G.}~\bibnamefont {Yamahata}}, \bibinfo
  {author} {\bibfnamefont {M.-H.}\ \bibnamefont {Bae}}, \bibinfo {author}
  {\bibfnamefont {N.}~\bibnamefont {Kim}}, \bibinfo {author} {\bibfnamefont
  {A.}~\bibnamefont {Rossi}}, \bibinfo {author} {\bibfnamefont
  {M.}~\bibnamefont {M{\"o}tt{\"o}nen}}, \ and\ \bibinfo {author}
  {\bibfnamefont {M.}~\bibnamefont {Kataoka}},\ }\bibfield  {title} {\enquote
  {\bibinfo {title} {Evidence for universality of tunable-barrier electron
  pumps},}\ }\href@noop {} {\bibfield  {journal} {\bibinfo  {journal}
  {Metrologia}\ }\textbf {\bibinfo {volume} {56}},\ \bibinfo {pages} {044004}
  (\bibinfo {year} {2019}{\natexlab{a}})}\BibitemShut {NoStop}%
\bibitem [{\citenamefont {Zhao}\ \emph {et~al.}(2017)\citenamefont {Zhao},
  \citenamefont {Rossi}, \citenamefont {Giblin}, \citenamefont {Fletcher},
  \citenamefont {Hudson}, \citenamefont {M{\"o}tt{\"o}nen}, \citenamefont
  {Kataoka},\ and\ \citenamefont {Dzurak}}]{zhao2017thermal}%
  \BibitemOpen
  \bibfield  {author} {\bibinfo {author} {\bibfnamefont {R.}~\bibnamefont
  {Zhao}}, \bibinfo {author} {\bibfnamefont {A.}~\bibnamefont {Rossi}},
  \bibinfo {author} {\bibfnamefont {S.}~\bibnamefont {Giblin}}, \bibinfo
  {author} {\bibfnamefont {J.}~\bibnamefont {Fletcher}}, \bibinfo {author}
  {\bibfnamefont {F.}~\bibnamefont {Hudson}}, \bibinfo {author} {\bibfnamefont
  {M.}~\bibnamefont {M{\"o}tt{\"o}nen}}, \bibinfo {author} {\bibfnamefont
  {M.}~\bibnamefont {Kataoka}}, \ and\ \bibinfo {author} {\bibfnamefont
  {A.}~\bibnamefont {Dzurak}},\ }\bibfield  {title} {\enquote {\bibinfo {title}
  {Thermal-error regime in high-accuracy gigahertz single-electron pumping},}\
  }\href@noop {} {\bibfield  {journal} {\bibinfo  {journal} {Physical Review
  Applied}\ }\textbf {\bibinfo {volume} {8}},\ \bibinfo {pages} {044021}
  (\bibinfo {year} {2017})}\BibitemShut {NoStop}%
\bibitem [{\citenamefont {Stein}\ \emph {et~al.}(2017)\citenamefont {Stein},
  \citenamefont {Scherer}, \citenamefont {Gerster}, \citenamefont {Behr},
  \citenamefont {G{\"o}tz}, \citenamefont {Pesel}, \citenamefont {Leicht},
  \citenamefont {Ubbelohde}, \citenamefont {Weimann}, \citenamefont {Pierz}
  \emph {et~al.}}]{stein2016robustness}%
  \BibitemOpen
  \bibfield  {author} {\bibinfo {author} {\bibfnamefont {F.}~\bibnamefont
  {Stein}}, \bibinfo {author} {\bibfnamefont {H.}~\bibnamefont {Scherer}},
  \bibinfo {author} {\bibfnamefont {T.}~\bibnamefont {Gerster}}, \bibinfo
  {author} {\bibfnamefont {R.}~\bibnamefont {Behr}}, \bibinfo {author}
  {\bibfnamefont {M.}~\bibnamefont {G{\"o}tz}}, \bibinfo {author}
  {\bibfnamefont {E.}~\bibnamefont {Pesel}}, \bibinfo {author} {\bibfnamefont
  {C.}~\bibnamefont {Leicht}}, \bibinfo {author} {\bibfnamefont
  {N.}~\bibnamefont {Ubbelohde}}, \bibinfo {author} {\bibfnamefont
  {T.}~\bibnamefont {Weimann}}, \bibinfo {author} {\bibfnamefont
  {K.}~\bibnamefont {Pierz}},  \emph {et~al.},\ }\bibfield  {title} {\enquote
  {\bibinfo {title} {Robustness of single-electron pumps at sub-ppm current
  accuracy level},}\ }\href@noop {} {\bibfield  {journal} {\bibinfo  {journal}
  {Metrologia}\ }\textbf {\bibinfo {volume} {54}},\ \bibinfo {pages} {S1}
  (\bibinfo {year} {2017})}\BibitemShut {NoStop}%
\bibitem [{\citenamefont {Yamahata}\ \emph {et~al.}(2016)\citenamefont
  {Yamahata}, \citenamefont {Giblin}, \citenamefont {Kataoka}, \citenamefont
  {Karasawa},\ and\ \citenamefont {Fujiwara}}]{yamahata2016gigahertz}%
  \BibitemOpen
  \bibfield  {author} {\bibinfo {author} {\bibfnamefont {G.}~\bibnamefont
  {Yamahata}}, \bibinfo {author} {\bibfnamefont {S.~P.}\ \bibnamefont
  {Giblin}}, \bibinfo {author} {\bibfnamefont {M.}~\bibnamefont {Kataoka}},
  \bibinfo {author} {\bibfnamefont {T.}~\bibnamefont {Karasawa}}, \ and\
  \bibinfo {author} {\bibfnamefont {A.}~\bibnamefont {Fujiwara}},\ }\bibfield
  {title} {\enquote {\bibinfo {title} {Gigahertz single-electron pumping in
  silicon with an accuracy better than 9.2 parts in 10$^{7}$},}\ }\href@noop {}
  {\bibfield  {journal} {\bibinfo  {journal} {Applied Physics Letters}\
  }\textbf {\bibinfo {volume} {109}},\ \bibinfo {pages} {013101} (\bibinfo
  {year} {2016})}\BibitemShut {NoStop}%
\bibitem [{\citenamefont {Fujiwara}, \citenamefont {Nishiguchi},\ and\
  \citenamefont {Ono}(2008)}]{fujiwara2008nanoampere}%
  \BibitemOpen
  \bibfield  {author} {\bibinfo {author} {\bibfnamefont {A.}~\bibnamefont
  {Fujiwara}}, \bibinfo {author} {\bibfnamefont {K.}~\bibnamefont
  {Nishiguchi}}, \ and\ \bibinfo {author} {\bibfnamefont {Y.}~\bibnamefont
  {Ono}},\ }\bibfield  {title} {\enquote {\bibinfo {title} {Nanoampere charge
  pump by single-electron ratchet using silicon nanowire
  metal-oxide-semiconductor field-effect transistor},}\ }\href@noop {}
  {\bibfield  {journal} {\bibinfo  {journal} {Applied Physics Letters}\
  }\textbf {\bibinfo {volume} {92}},\ \bibinfo {pages} {042102} (\bibinfo
  {year} {2008})}\BibitemShut {NoStop}%
\bibitem [{\citenamefont {Drung}\ \emph {et~al.}(2015)\citenamefont {Drung},
  \citenamefont {Krause}, \citenamefont {Becker}, \citenamefont {Scherer},\
  and\ \citenamefont {Ahlers}}]{drung2015ultrastable}%
  \BibitemOpen
  \bibfield  {author} {\bibinfo {author} {\bibfnamefont {D.}~\bibnamefont
  {Drung}}, \bibinfo {author} {\bibfnamefont {C.}~\bibnamefont {Krause}},
  \bibinfo {author} {\bibfnamefont {U.}~\bibnamefont {Becker}}, \bibinfo
  {author} {\bibfnamefont {H.}~\bibnamefont {Scherer}}, \ and\ \bibinfo
  {author} {\bibfnamefont {F.~J.}\ \bibnamefont {Ahlers}},\ }\bibfield  {title}
  {\enquote {\bibinfo {title} {Ultrastable low-noise current amplifier: A novel
  device for measuring small electric currents with high accuracy},}\
  }\href@noop {} {\bibfield  {journal} {\bibinfo  {journal} {Review of
  Scientific Instruments}\ }\textbf {\bibinfo {volume} {86}},\ \bibinfo {pages}
  {024703} (\bibinfo {year} {2015})}\BibitemShut {NoStop}%
\bibitem [{\citenamefont {Krause}\ \emph {et~al.}(2019)\citenamefont {Krause},
  \citenamefont {Drung}, \citenamefont {G{\"o}tz},\ and\ \citenamefont
  {Scherer}}]{krause2019noise}%
  \BibitemOpen
  \bibfield  {author} {\bibinfo {author} {\bibfnamefont {C.}~\bibnamefont
  {Krause}}, \bibinfo {author} {\bibfnamefont {D.}~\bibnamefont {Drung}},
  \bibinfo {author} {\bibfnamefont {M.}~\bibnamefont {G{\"o}tz}}, \ and\
  \bibinfo {author} {\bibfnamefont {H.}~\bibnamefont {Scherer}},\ }\bibfield
  {title} {\enquote {\bibinfo {title} {Noise-optimized ultrastable low-noise
  current amplifier},}\ }\href@noop {} {\bibfield  {journal} {\bibinfo
  {journal} {Review of Scientific Instruments}\ }\textbf {\bibinfo {volume}
  {90}},\ \bibinfo {pages} {014706} (\bibinfo {year} {2019})}\BibitemShut
  {NoStop}%
\bibitem [{\citenamefont {Giblin}\ \emph
  {et~al.}(2019{\natexlab{b}})\citenamefont {Giblin}, \citenamefont {Drung},
  \citenamefont {G{\"o}tz},\ and\ \citenamefont
  {Scherer}}]{giblin2019interlaboratory}%
  \BibitemOpen
  \bibfield  {author} {\bibinfo {author} {\bibfnamefont {S.~P.}\ \bibnamefont
  {Giblin}}, \bibinfo {author} {\bibfnamefont {D.}~\bibnamefont {Drung}},
  \bibinfo {author} {\bibfnamefont {M.}~\bibnamefont {G{\"o}tz}}, \ and\
  \bibinfo {author} {\bibfnamefont {H.}~\bibnamefont {Scherer}},\ }\bibfield
  {title} {\enquote {\bibinfo {title} {Interlaboratory nanoamp current
  comparison with subpart-per-million uncertainty},}\ }\href {\doibase
  10.1109/TIM.2018.2879126} {\bibfield  {journal} {\bibinfo  {journal} {IEEE
  Transactions on Instrumentation and Measurement}\ }\textbf {\bibinfo {volume}
  {68}},\ \bibinfo {pages} {1996--2002} (\bibinfo {year}
  {2019}{\natexlab{b}})}\BibitemShut {NoStop}%
\bibitem [{Note1()}]{Note1}%
  \BibitemOpen
  \bibinfo {note} {Mention of specific models of commercial instrument is for
  information only and does not imply endorsement by the authors or their
  respective institutions}\BibitemShut {NoStop}%
\bibitem [{\citenamefont {Kaestner}\ \emph {et~al.}(2008)\citenamefont
  {Kaestner}, \citenamefont {Kashcheyevs}, \citenamefont {Hein}, \citenamefont
  {Pierz}, \citenamefont {Siegner},\ and\ \citenamefont
  {Schumacher}}]{kaestner2008robust}%
  \BibitemOpen
  \bibfield  {author} {\bibinfo {author} {\bibfnamefont {B.}~\bibnamefont
  {Kaestner}}, \bibinfo {author} {\bibfnamefont {V.}~\bibnamefont
  {Kashcheyevs}}, \bibinfo {author} {\bibfnamefont {G.}~\bibnamefont {Hein}},
  \bibinfo {author} {\bibfnamefont {K.}~\bibnamefont {Pierz}}, \bibinfo
  {author} {\bibfnamefont {U.}~\bibnamefont {Siegner}}, \ and\ \bibinfo
  {author} {\bibfnamefont {H.~W.}\ \bibnamefont {Schumacher}},\ }\bibfield
  {title} {\enquote {\bibinfo {title} {Robust single-parameter quantized charge
  pumping},}\ }\href@noop {} {\bibfield  {journal} {\bibinfo  {journal}
  {Applied Physics Letters}\ }\textbf {\bibinfo {volume} {92}},\ \bibinfo
  {pages} {192106} (\bibinfo {year} {2008})}\BibitemShut {NoStop}%
\bibitem [{\citenamefont {Giblin}\ \emph {et~al.}(2017)\citenamefont {Giblin},
  \citenamefont {Bae}, \citenamefont {Kim}, \citenamefont {Ahn},\ and\
  \citenamefont {Kataoka}}]{giblin2017robust}%
  \BibitemOpen
  \bibfield  {author} {\bibinfo {author} {\bibfnamefont {S.}~\bibnamefont
  {Giblin}}, \bibinfo {author} {\bibfnamefont {M.}~\bibnamefont {Bae}},
  \bibinfo {author} {\bibfnamefont {N.}~\bibnamefont {Kim}}, \bibinfo {author}
  {\bibfnamefont {Y.-H.}\ \bibnamefont {Ahn}}, \ and\ \bibinfo {author}
  {\bibfnamefont {M.}~\bibnamefont {Kataoka}},\ }\bibfield  {title} {\enquote
  {\bibinfo {title} {Robust operation of a gallium arsenide tunable barrier
  electron pump},}\ }\href@noop {} {\bibfield  {journal} {\bibinfo  {journal}
  {Metrologia}\ }\textbf {\bibinfo {volume} {54}},\ \bibinfo {pages} {299}
  (\bibinfo {year} {2017})}\BibitemShut {NoStop}%
\bibitem [{\citenamefont {Delahaye}\ and\ \citenamefont
  {Jeckelmann}(2003)}]{delahaye2003revised}%
  \BibitemOpen
  \bibfield  {author} {\bibinfo {author} {\bibfnamefont {F.}~\bibnamefont
  {Delahaye}}\ and\ \bibinfo {author} {\bibfnamefont {B.}~\bibnamefont
  {Jeckelmann}},\ }\bibfield  {title} {\enquote {\bibinfo {title} {{Revised
  technical guidelines for reliable dc measurements of the quantized Hall
  resistance}},}\ }\href@noop {} {\bibfield  {journal} {\bibinfo  {journal}
  {Metrologia}\ }\textbf {\bibinfo {volume} {40}},\ \bibinfo {pages} {217}
  (\bibinfo {year} {2003})}\BibitemShut {NoStop}%
\bibitem [{\citenamefont {Zimmerman}\ \emph {et~al.}(2001)\citenamefont
  {Zimmerman}, \citenamefont {Huber}, \citenamefont {Fujiwara},\ and\
  \citenamefont {Takahashi}}]{zimmerman2001excellent}%
  \BibitemOpen
  \bibfield  {author} {\bibinfo {author} {\bibfnamefont {N.~M.}\ \bibnamefont
  {Zimmerman}}, \bibinfo {author} {\bibfnamefont {W.~H.}\ \bibnamefont
  {Huber}}, \bibinfo {author} {\bibfnamefont {A.}~\bibnamefont {Fujiwara}}, \
  and\ \bibinfo {author} {\bibfnamefont {Y.}~\bibnamefont {Takahashi}},\
  }\bibfield  {title} {\enquote {\bibinfo {title} {Excellent charge offset
  stability in a si-based single-electron tunneling transistor},}\ }\href@noop
  {} {\bibfield  {journal} {\bibinfo  {journal} {Applied Physics Letters}\
  }\textbf {\bibinfo {volume} {79}},\ \bibinfo {pages} {3188--3190} (\bibinfo
  {year} {2001})}\BibitemShut {NoStop}%
\bibitem [{\citenamefont {Hourdakis}, \citenamefont {Wahl},\ and\ \citenamefont
  {Zimmerman}(2008)}]{hourdakis2008lack}%
  \BibitemOpen
  \bibfield  {author} {\bibinfo {author} {\bibfnamefont {E.}~\bibnamefont
  {Hourdakis}}, \bibinfo {author} {\bibfnamefont {J.~A.}\ \bibnamefont {Wahl}},
  \ and\ \bibinfo {author} {\bibfnamefont {N.~M.}\ \bibnamefont {Zimmerman}},\
  }\bibfield  {title} {\enquote {\bibinfo {title} {Lack of charge offset drift
  is a robust property of si single electron transistors},}\ }\href@noop {}
  {\bibfield  {journal} {\bibinfo  {journal} {Applied Physics Letters}\
  }\textbf {\bibinfo {volume} {92}},\ \bibinfo {pages} {062102} (\bibinfo
  {year} {2008})}\BibitemShut {NoStop}%
\bibitem [{\citenamefont {Yamahata}\ \emph {et~al.}(2019)\citenamefont
  {Yamahata}, \citenamefont {Ryu}, \citenamefont {Johnson}, \citenamefont
  {Sim}, \citenamefont {Fujiwara},\ and\ \citenamefont
  {Kataoka}}]{yamahata2019picosecond}%
  \BibitemOpen
  \bibfield  {author} {\bibinfo {author} {\bibfnamefont {G.}~\bibnamefont
  {Yamahata}}, \bibinfo {author} {\bibfnamefont {S.}~\bibnamefont {Ryu}},
  \bibinfo {author} {\bibfnamefont {N.}~\bibnamefont {Johnson}}, \bibinfo
  {author} {\bibfnamefont {H.-S.}\ \bibnamefont {Sim}}, \bibinfo {author}
  {\bibfnamefont {A.}~\bibnamefont {Fujiwara}}, \ and\ \bibinfo {author}
  {\bibfnamefont {M.}~\bibnamefont {Kataoka}},\ }\bibfield  {title} {\enquote
  {\bibinfo {title} {Picosecond coherent electron motion in a silicon
  single-electron source},}\ }\href@noop {} {\bibfield  {journal} {\bibinfo
  {journal} {Nature nanotechnology}\ }\textbf {\bibinfo {volume} {14}},\
  \bibinfo {pages} {1019--1023} (\bibinfo {year} {2019})}\BibitemShut {NoStop}%
\bibitem [{\citenamefont {Gerster}\ \emph {et~al.}(2019)\citenamefont
  {Gerster}, \citenamefont {M{\"u}ller}, \citenamefont {Freise}, \citenamefont
  {Reifert}, \citenamefont {Maradan}, \citenamefont {Hinze}, \citenamefont
  {Weimann}, \citenamefont {Marx}, \citenamefont {Pierz}, \citenamefont
  {Schumacher} \emph {et~al.}}]{gerster2019robust}%
  \BibitemOpen
  \bibfield  {author} {\bibinfo {author} {\bibfnamefont {T.}~\bibnamefont
  {Gerster}}, \bibinfo {author} {\bibfnamefont {A.}~\bibnamefont {M{\"u}ller}},
  \bibinfo {author} {\bibfnamefont {L.}~\bibnamefont {Freise}}, \bibinfo
  {author} {\bibfnamefont {D.}~\bibnamefont {Reifert}}, \bibinfo {author}
  {\bibfnamefont {D.}~\bibnamefont {Maradan}}, \bibinfo {author} {\bibfnamefont
  {P.}~\bibnamefont {Hinze}}, \bibinfo {author} {\bibfnamefont
  {T.}~\bibnamefont {Weimann}}, \bibinfo {author} {\bibfnamefont
  {H.}~\bibnamefont {Marx}}, \bibinfo {author} {\bibfnamefont {K.}~\bibnamefont
  {Pierz}}, \bibinfo {author} {\bibfnamefont {H.}~\bibnamefont {Schumacher}},
  \emph {et~al.},\ }\bibfield  {title} {\enquote {\bibinfo {title} {Robust
  formation of quantum dots in gaas/algaas heterostructures for single-electron
  metrology},}\ }\href@noop {} {\bibfield  {journal} {\bibinfo  {journal}
  {Metrologia}\ }\textbf {\bibinfo {volume} {56}},\ \bibinfo {pages} {014002}
  (\bibinfo {year} {2019})}\BibitemShut {NoStop}%
\bibitem [{\citenamefont {Stein}\ \emph {et~al.}(2015)\citenamefont {Stein},
  \citenamefont {Drung}, \citenamefont {Fricke}, \citenamefont {Scherer},
  \citenamefont {Hohls}, \citenamefont {Leicht}, \citenamefont {Goetz},
  \citenamefont {Krause}, \citenamefont {Behr}, \citenamefont {Pesel},
  \citenamefont {Siegner}, \citenamefont {Ahlers},\ and\ \citenamefont
  {Schumacher}}]{stein2015validation}%
  \BibitemOpen
  \bibfield  {author} {\bibinfo {author} {\bibfnamefont {F.}~\bibnamefont
  {Stein}}, \bibinfo {author} {\bibfnamefont {D.}~\bibnamefont {Drung}},
  \bibinfo {author} {\bibfnamefont {L.}~\bibnamefont {Fricke}}, \bibinfo
  {author} {\bibfnamefont {H.}~\bibnamefont {Scherer}}, \bibinfo {author}
  {\bibfnamefont {F.}~\bibnamefont {Hohls}}, \bibinfo {author} {\bibfnamefont
  {C.}~\bibnamefont {Leicht}}, \bibinfo {author} {\bibfnamefont
  {M.}~\bibnamefont {Goetz}}, \bibinfo {author} {\bibfnamefont
  {C.}~\bibnamefont {Krause}}, \bibinfo {author} {\bibfnamefont
  {R.}~\bibnamefont {Behr}}, \bibinfo {author} {\bibfnamefont {E.}~\bibnamefont
  {Pesel}}, \bibinfo {author} {\bibfnamefont {U.}~\bibnamefont {Siegner}},
  \bibinfo {author} {\bibfnamefont {F.-J.}\ \bibnamefont {Ahlers}}, \ and\
  \bibinfo {author} {\bibfnamefont {H.~W.}\ \bibnamefont {Schumacher}},\
  }\bibfield  {title} {\enquote {\bibinfo {title} {validation of a
  quantized-current source with 0.2 ppm uncertainty},}\ }\href@noop {}
  {\bibfield  {journal} {\bibinfo  {journal} {Applied Physics Letters}\
  }\textbf {\bibinfo {volume} {107}},\ \bibinfo {pages} {103501} (\bibinfo
  {year} {2015})}\BibitemShut {NoStop}%
\bibitem [{\citenamefont {Giblin}\ \emph {et~al.}(2012)\citenamefont {Giblin},
  \citenamefont {Kataoka}, \citenamefont {Fletcher}, \citenamefont {See},
  \citenamefont {Janssen}, \citenamefont {Griffiths}, \citenamefont {Jones},
  \citenamefont {Farrer},\ and\ \citenamefont {Ritchie}}]{giblin2012towards}%
  \BibitemOpen
  \bibfield  {author} {\bibinfo {author} {\bibfnamefont {S.~P.}\ \bibnamefont
  {Giblin}}, \bibinfo {author} {\bibfnamefont {M.}~\bibnamefont {Kataoka}},
  \bibinfo {author} {\bibfnamefont {J.~D.}\ \bibnamefont {Fletcher}}, \bibinfo
  {author} {\bibfnamefont {P.}~\bibnamefont {See}}, \bibinfo {author}
  {\bibfnamefont {T.}~\bibnamefont {Janssen}}, \bibinfo {author} {\bibfnamefont
  {J.~P.}\ \bibnamefont {Griffiths}}, \bibinfo {author} {\bibfnamefont
  {G.~A.~C.}\ \bibnamefont {Jones}}, \bibinfo {author} {\bibfnamefont
  {I.}~\bibnamefont {Farrer}}, \ and\ \bibinfo {author} {\bibfnamefont {D.~A.}\
  \bibnamefont {Ritchie}},\ }\bibfield  {title} {\enquote {\bibinfo {title}
  {Towards a quantum representation of the ampere using single electron
  pumps},}\ }\href@noop {} {\bibfield  {journal} {\bibinfo  {journal} {Nature
  Communications}\ }\textbf {\bibinfo {volume} {3}},\ \bibinfo {pages} {930}
  (\bibinfo {year} {2012})}\BibitemShut {NoStop}%
\bibitem [{\citenamefont {Bae}\ \emph {et~al.}(2015)\citenamefont {Bae},
  \citenamefont {Ahn}, \citenamefont {Seo}, \citenamefont {Chung},
  \citenamefont {Fletcher}, \citenamefont {Giblin}, \citenamefont {Kataoka},\
  and\ \citenamefont {Kim}}]{bae2015precision}%
  \BibitemOpen
  \bibfield  {author} {\bibinfo {author} {\bibfnamefont {M.-H.}\ \bibnamefont
  {Bae}}, \bibinfo {author} {\bibfnamefont {Y.-H.}\ \bibnamefont {Ahn}},
  \bibinfo {author} {\bibfnamefont {M.}~\bibnamefont {Seo}}, \bibinfo {author}
  {\bibfnamefont {Y.}~\bibnamefont {Chung}}, \bibinfo {author} {\bibfnamefont
  {J.~D.}\ \bibnamefont {Fletcher}}, \bibinfo {author} {\bibfnamefont {S.~P.}\
  \bibnamefont {Giblin}}, \bibinfo {author} {\bibfnamefont {M.}~\bibnamefont
  {Kataoka}}, \ and\ \bibinfo {author} {\bibfnamefont {N.}~\bibnamefont
  {Kim}},\ }\bibfield  {title} {\enquote {\bibinfo {title} {Precision
  measurement of a potential-profile tunable single-electron pump},}\
  }\href@noop {} {\bibfield  {journal} {\bibinfo  {journal} {Metrologia}\
  }\textbf {\bibinfo {volume} {52}},\ \bibinfo {pages} {195} (\bibinfo {year}
  {2015})}\BibitemShut {NoStop}%
\end{thebibliography}%
\newpage
\section{Supplementary Information}

\setcounter{figure}{0}
\renewcommand{\thefigure}{S\arabic{figure}}

\subsection{Raw data and noise}

\begin{figure}
\includegraphics[width=9cm]{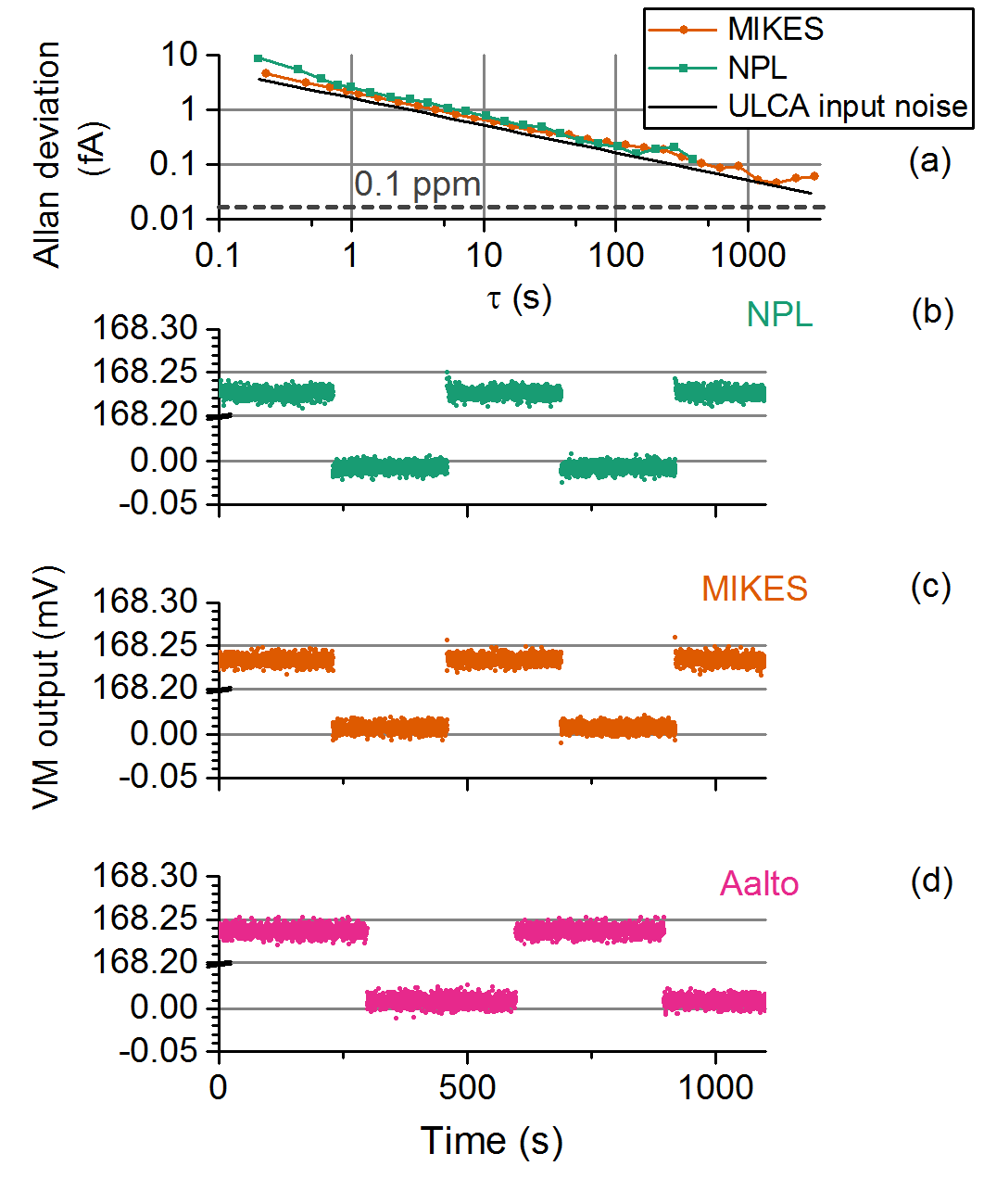}
\caption{\label{RawDataFig}\textsf{(a): Filled symbols: Allan deviation computed from time-domain data traces at NPL and MIKES, in which the pump was continuously on. Solid line: theoretical Allan deviation for the Johnson current noise due to the ULCA input resistance. (b)-(d): Raw DVM data, recording the ULCA output for precision measurements at NPL, MIKES and Aalto. In these measurements, the on-off cycle is effected by turning the RF pump signal on and off.}}
\end{figure}

Long integration times are required to resolve small currents to high resolution, making it important to account for the different sources of noise in the experiment. In figure \ref{RawDataFig}(a), we show the Allan deviation of the current with the pump tuned to the $N=1$ plateau and turned on, at both NPL and MIKES. We compare these to the Allan deviation expected for the Johnson current noise in the ULCA input. This noise is expected to be the largest single contribution to the total noise. It has a spectral density per unit bandwidth of $\sqrt{4k_{\text{B}}T/R}$, with $R=3$~G$\Omega$: approximately $2.3$~fA/$\sqrt{\text{Hz}}$. The experimental Allan deviation is marginally above the ULCA noise floor due to small additional contributions such as the ULCA output noise and additional noise due to vibrations in the wiring of the cryogenic probe. The excellent stability of the ULCA\cite{drung2015ultrastable} is clear, particularly from the MIKES data trace where there is no clear transition to a regime of $1/f$ noise for averaging times as long as $1000$~s. This stability permits a long timescale for the on-off cycle of the pump measurement, a key advantage in the presence of unexpected time constants as discussed in section D below.

In figures \ref{RawDataFig}(b-d) we show sections of raw time-domain data from high-precision pump measurements in which the AC drive to the pump was turned on and off. The presented data are segments of the data sets from runs NPL1, MIKES1 and Aalto shown in figure \ref{SummaryHighAccFig} of the main text. The only quantitative difference visible by eye is a slight change in the offset in the `pump off' state between the NPL and MIKES systems. The differences in the ULCA gain and voltmeter calibration factors between the NPL and MIKES setups give a difference in the indicated on-off voltage difference of $\sim 40$~ppm, or around $7$~$\mu$V.

\subsection{ULCA stability}

\begin{figure}
\includegraphics[width=9cm]{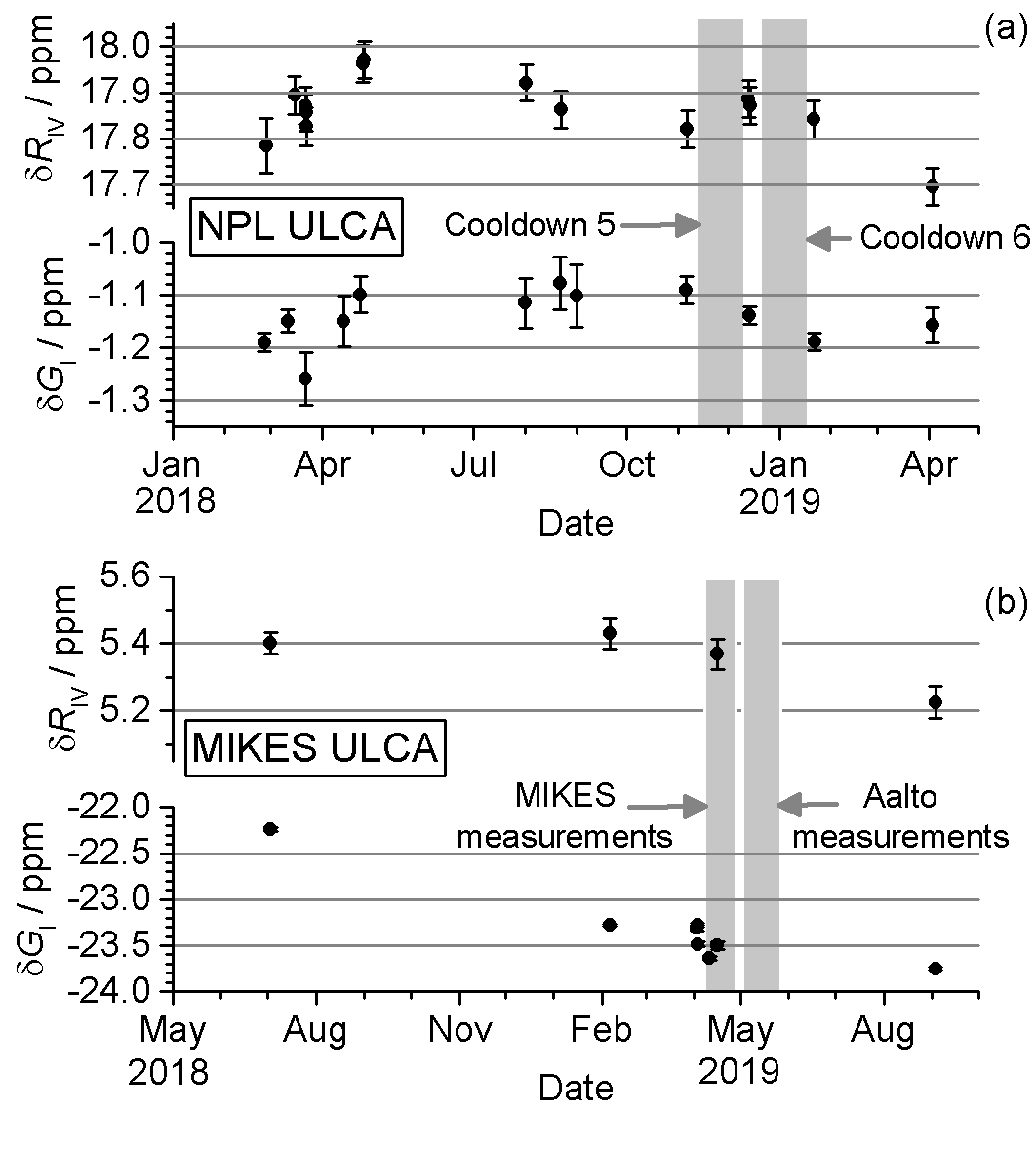}
\caption{\label{ULCACalFig}\textsf{Deviations from the nominal value of the output transresistance gain $R_{\text{IV}}$ and of the input current gain $G_{\text{I}}$ for (a) the NPL ULCA and (b) the MIKES ULCA. The time-spans of high-precision electron pump measurement campaigns are indicated by vertical grey bars.}}
\end{figure}

In previous high-precision measurements of electron pump currents, the stability of the measuring system, whether based on a $1$~G$\Omega$ standard resistor\cite{zhao2017thermal} or an ULCA\cite{stein2016robustness}, has become an important contribution to the overall uncertainty. In this work, an ULCA was used to convert the pump current to a voltage for subsequent readout by a precision-calibrated DVM. In figure \ref{ULCACalFig}, we present the calibration data for the two ULCA units employed: the NPL ULCA at NPL, and the MIKES ULCA at MIKES and Aalto. The time intervals for the high-precision electron pump measurements reported in the paper are indicated by vertical grey bands on both plots. Both ULCAs demonstrate a gain stability consistent with a published study, in which several ULCA units demonstrated a drift in the overall transresistance gain less than $1$~ppm per year\cite{krause2019noise}. The change in the ULCA gain from one calibration to the next justifies a drift uncertainty term (see table 1 of the main text) of less than 0.1 ppm when a rectangular distribution is assumed to describe the gain in the time interval between calibrations. This drift uncertainty was evaluated using the rectangular distribution in the same way as the drift uncertainty of the DVM calibration factor detailed in the next section, and in the supplementary information for Ref. \onlinecite{zhao2017thermal}

\subsection{NPL high-precision measurement setup}

\begin{figure}
\includegraphics[width=9cm]{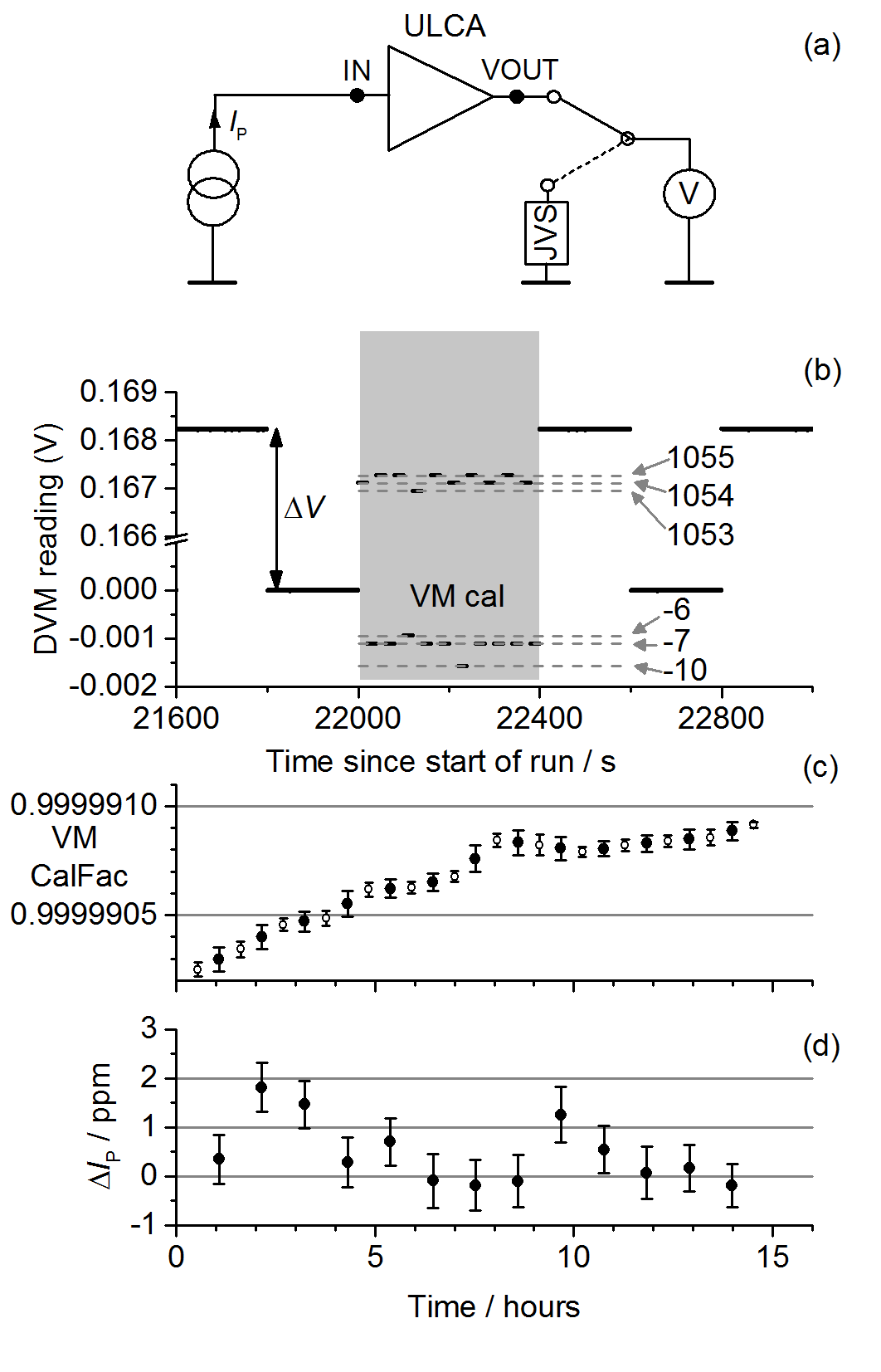}
\caption{\label{InterleavedDataFig}\textsf{(a): Schematic diagram of the NPL high-precision measurement circuit. JVS = Josephson Voltage Standard. (b): Section of raw voltmeter data taken from run NPL 1. The Josephson voltage step numbers during the DVM calibration cycles are indicated on the plot. A vertical arrow shows the difference voltage $\Delta V$ extracted from the pump measurement cycle. (c): DVM calibration factors measured during run NPL 2. Open circles: Measured DVM calibration factor. Filled circles: average of the adjacent calibration factors used for analysing each sequence of pump measurement. (d): Pump current evaluated from the 13 pump measurement sequences of run NPL 2. Plots (c) and (d) share the same x-axis.}}
\end{figure}

In all the reported high-precision measurements, the ULCA output was recorded by a precision digital voltmeter (DVM), the Keysight 3458A. The DVM is calibrated to yield the calibration factor $C$ defined as the true voltage divided by the indicated voltage. Exceptionally stable examples of this model of DVM have demonstrated stability of the calibration factor as good as a few parts in $10^{8}$ per day \cite{stein2015validation}. However, our experience in this study, and with previous studies at NPL\cite{yamahata2016gigahertz,giblin2017robust,zhao2017thermal} is that the $C$ could change by up to a few parts in $10^7$ per day. Even with daily calibrations directly against a Josephson array, as in the MIKES measurements, the use of a rectangular distribution yielded a typical relative uncertainty due to drift of around $10^{-7}$.  To lower this uncertainty contribution, at NPL the DVM calibration interval was reduced to an hour by incorporating a switch (using one channel of a Data Proof DP320 scanner) into the measurement system as shown in figure \ref{InterleavedDataFig}(a). This switch allowed voltmeter calibrations to be interleaved with pump measurements, enabling a dramatic reduction in the uncertainty due to the voltmeter calibration. Figure \ref{InterleavedDataFig}(b) shows a section of raw DVM data from measurement NPL 1. On the left-hand side of the plot for elapsed run-time less than 22000 s, the input of the DVM is connected to the pump, and an on-off cycle is visible with $1000$ data points for each on and off segment. Then the switch connects the DVM to the Josephson array and $10$ on-off cycles are executed with $100$ data points per segment. A shorter cycle is possible during the calibration step, because the time constant is negligible. Because the Josephson array is a hysteretic type operating at zero current bias, the same voltage step is not obtained each time the array voltage is changed. However, this is not important since we are only interested in the gain factor of the DVM, not the offset. The array voltage is given by $V_{\text{J}} = nf/K_{\text{J-90}}$, with the step numbers $n$ indicated on the plot and $f_{\text{J}} = 76.674$~GHz. Note that the calibration voltage and the ULCA output during a pump measurement are roughly equal, $\sim 168$~mV, and hence there is no need to correct for or add uncertainties due to the DVM non-linearity. The uncertainty in the DVM calibration factor from $10$ calibration on-off cycles is typically in the range from $3 \times 10^{-8}$ to $5 \times 10^{-8}$. The drift over the 1-hour calibration interval is also typically less than $5 \times 10^{-8}$ although interestingly, larger jumps are occasionally observed with the largest jumps being a few parts in $10^{-7}$, similar to the typical jumps in a 24 hour calibration interval. Each set of $10$ pump measurement cycles is analyzed using the mean of the DVM calibration factors before and after the set, and the uncertainty in the DVM calibration is added to the statistical uncertainty of the pump measurement.  The precision measurement run `NPL 1' consists of $16$ sets of $10$ pump cycles each, interleaved with sets of $10$ DVM calibration cycles, and lasts $22$ hours.

Next we discuss the process of analysing the interleaved data in more detail, with reference to figure \ref{InterleavedDataFig} (c) and (d). In figure \ref{InterleavedDataFig} (c), we show the DVM calibration factors from run NPL 2. The error bars are the combination of type A and type B uncertainties in the DVM calibration, and are dominated by the type A contribution. Each sequence of pump measurement cycles is analyzed using the mean of the DVM calibration factors before and after the sequence: $C_{i} = (C_{\text{before}} + C_{\text{after}})/2$. These mean values are shown as filled circles in figure \ref{InterleavedDataFig} (c). The uncertainty in the mean values (shown as the error bars on the plot) is the root-sum-square of three contributions: the uncertainties in the two calibrations before and after the sequence, and a drift uncertainty which is taken from the rectangular distribution as $U = |U_{\text{before}} - U_{\text{after}}|/(2\sqrt{3})$. In the analysis of the pump measurement sequence, we extract the difference signal $\Delta V$, and the pump current in the \textit{i}'th sequence is evaluated as $I_{\text{P}i} = C_{i} \Delta V / A_{\text{TR}}$, where $A_{\text{TR}}$ is the trans-resistance gain of the UCLA\cite{drung2015ultrastable}. The values of $\Delta I_{\text{P}i}$ are shown in figure \ref{InterleavedDataFig} (d). The error bars are the root-sum-square combination of the type A uncertainty in $\Delta V$, and the uncertainty in $C$. The weighted mean of the 13 values of $\Delta I_{\text{Pi}}$ yields a normalised deviation of the pump current from $e_{\text{90}}f$ of $0.464 \pm 0.14 \times 10^{-6}$. Because this data was obtained by the gate-switching method, a leakage current correction is applied to obtain the value of $\Delta I_{\text{P}}$ shown in figure \ref{SummaryHighAccFig}.

An logical improvement to the setup would be to use the Josephson array voltage continuously as a voltage reference, so that the DVM measured a small difference between the ULCA output and $V_{\text{J}}$. This `null-voltage' approach was used in Ref. \onlinecite{stein2016robustness} and reduced the uncertainty due to the voltage calibration to a negligible level. In fact, it was attempted in the present study at NPL, but the JVS operation proved unreliable. This was probably due the fact that the JVS was in a separate laboratory to the electron pump, and the two were connected by a $\sim30$-m-long cable. This cable did not add any significant noise when it was connected to the high-impedance input of the DVM, but the null-voltage circuit inevitably requires the low-voltage sides of the pump experiment and the JVS to be connected. It is likely that in this configuration, circulating ground currents caused an unacceptable amount of current noise through the Josephson junctions and prevented stable operation of the JVS.

\begin{figure*}
\includegraphics[width=18cm]{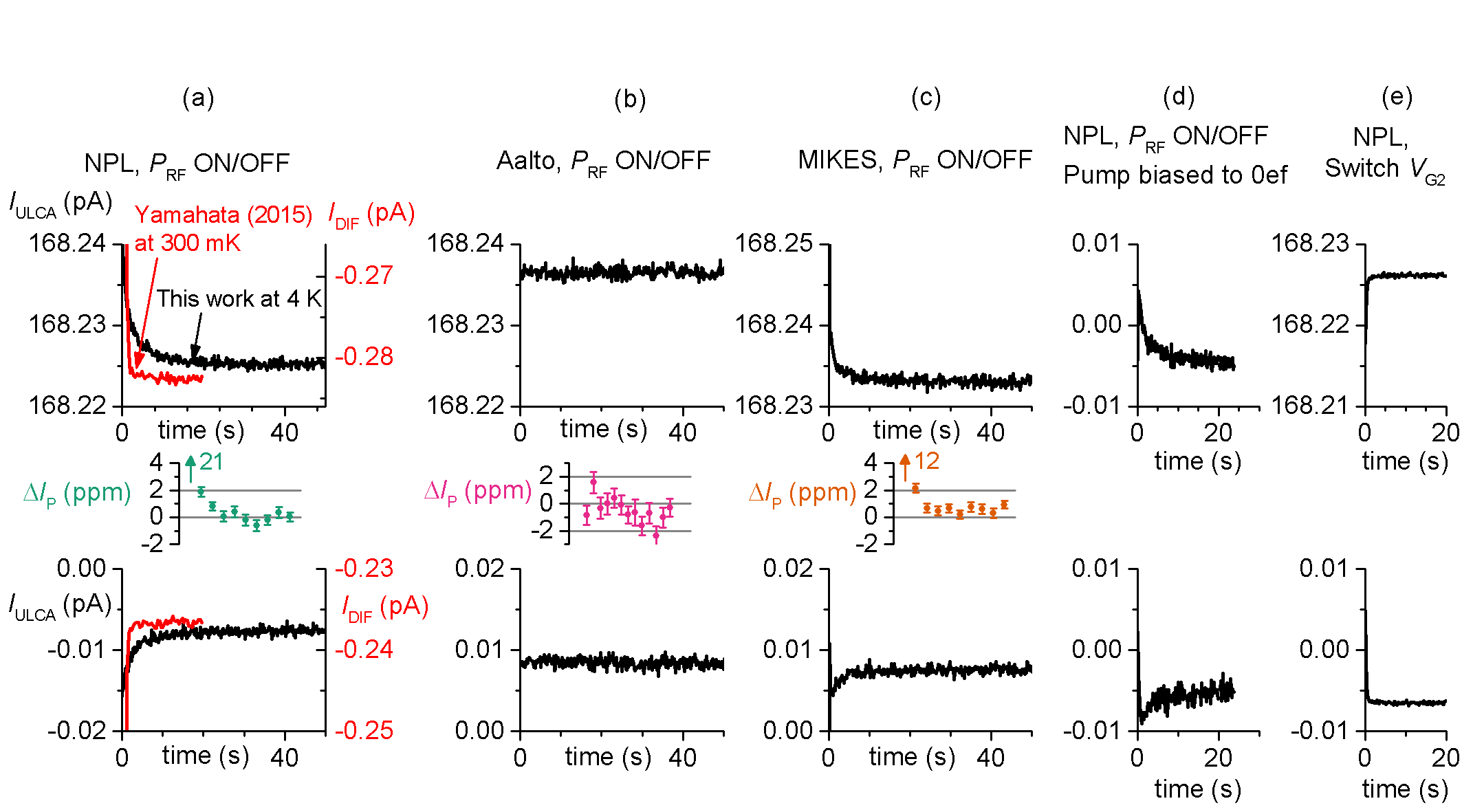}
\caption{\label{TimeConstantsFig}\textsf{Each main panel shows a point-by-point average (described in the text) of raw on-off data, with the y-axis of all plots covering the same range, $20$ fA. The top panels show the averages in the on state, and the bottom panels show the averages in the off state. (a): NPL measurements from this study using the power switching method (black line, left y-axis), and data from Ref. \onlinecite{yamahata2016gigahertz} for comparison (Red line, right y-axis). The measurements of Ref. \onlinecite{yamahata2016gigahertz} were obtained using a difference measurement system in which the raw current signal $I_{\text{DIF}}$ is the difference between the pump current and a reference current. (b): Aalto measurements using the power switching method. (c): MIKES measurements using the power switching method. (d): NPL measurement using the power switching method in which the pump was biased to the $N=0$ plateau. (e): NPL measurements using the gate stepping method. The lower apparent noise in this data is because a shorter cycle was used for the gate stepping on-off method (100 points instead of 1000 points per segment) and hence the point-by-point average is over a larger number of cycles than for the power switching method. The smaller inset panels in between the top and bottom main panels of (a), (b) and (c) show a 100-point moving-window analysis (described in the text) of the NPL, Aalto and MIKES precision measurements respectively, with vertical arrows indicating data points off the scale of the plot.}}
\end{figure*}

\subsection{Time constants}

When the RF drive signal is turned on or off, the pump current settles with a time constant which has been a few seconds in all previous precision studies at NPL\cite{giblin2012towards, bae2015precision,yamahata2016gigahertz,giblin2017robust,zhao2017thermal} and PTB\cite{stein2015validation,stein2016robustness}. In the present study, substantially longer time constants were observed. This required a correspondingly longer portion of the data to be rejected before analysis, and longer on-off cycles were used. Fortunately, the stability of the ULCA permitted on-off cycle times of several hundred seconds without $1/f$ noise compromising the type A uncertainty. In this supplementary section, we present data illustrating the time constants, and we discuss their possible origin.

Figure \ref{TimeConstantsFig} shows a temporal point-by-point averages of a number of representative raw sets of data. In this type of averaging, the first data point (at zero time) is the average of the first data points of all the `on' cycles (top panels) and all the `off' cycles (bottom panels). The second data point is the average of all the second data points, and so on. In panels (a), (b), and (c), we show the point-by-point averages of data measured at NPL, Aalto and MIKES in which the RF pump signal was turned on and off. For comparison, in panel (a) we have included data from Ref. \onlinecite{yamahata2016gigahertz}: there is clearly a time constant in the $4$~K measurements which was not present in the $1.5$~K measurements on the same sample. Different cryogenic probes were used in the two measurements, and the sample was thermalised differently. In the experiments of Ref. \onlinecite{yamahata2016gigahertz}, the sample was in vacuum and thermalised through the leads. In the present work, it was either immersed in liquid helium, or just above the liquid surface in helium vapour. The time constant is equal with the sample in liquid or vapour. The time constant observed in the NPL measurements is also present at MIKES but is completely absent in the Aalto data. For a different view of the time constant, the small inset panels show moving-window analyses of the precision measurements at NPL, MIKES and Aalto. In this analysis, $\Delta I_{\text{P}}$ is calculated from a window of 100 data points in each data segment, offset from the start of the segment by a multiple of 100 points. The first data point along the x-axis in each inset plot is calculated from points 1 to 100 in each raw data segment, the second data point from points 101 to 200, and so on up to the last point which is calculated from points 1201 to 1300 for the Aalto data, and points 901 to 1000 for the NPL and MIKES data. The time constants in the NPL and MIKES measurements, and the absence of a time constant in the Aalto measurement, are also visible in this analysis. The $\sim 0.5$~ppm offset in the pump current in the MIKES measurements is clearly resolved, even with this analysis which rejects $90 \%$ of the data.

Figure \ref{TimeConstantsFig}(d) shows a point-by-point average from a test measurement in which the RF power is turned on and off, but with the pump biased off the plateau with $V_{\text{EXIT}}=-1.7$~V, so that $I_{\text{P}}=0$. Turning the RF power on and off yields a transient current with a time constant very similar to that of the pumping data. The size of this transient is strongly dependent on $P_{\text{RF}}$, and disappears for $P_{\text{RF}}<8$~dBm. This suggests that the transient current is due to heating caused by the dissipation of RF power in the vicinity of the sample. Figure \ref{TimeConstantsFig}(e) shows data for the gate-stepping cycle, in which the RF drive was on all the time, $V_{\text{EXIT}}$ was switched from its operating point to a more negative value to turn the pump off. Here, the time constant is much shorter. 

We briefly speculate on the origin of the time constant using the power-switching method. One possible mechanism is temperature dependence of the leakage current which flows between the gate voltage lead and the pump channel, and contributes to the measured current. This leakage current is the parallel sum of several contributions due to the experimental wiring, sample holder circuit board, and possibly the sample gate oxide itself, and it is likely to be temperature dependent. Temperature changes of the relevant components due to heating from the RF power possibly result in a thermal time constant appearing in the current. This hypothesis is supported by the data of panel (d), in which the time constant appears in the absence of any pump current, with the pump tuned far from the $N=1$ plateau. However, the total absence of a time constant in the Aalto data of Figure \ref{TimeConstantsFig}(b) remains without a convincing explanation.

\subsection{Leakage correction}

\begin{figure}
\includegraphics[width=9cm]{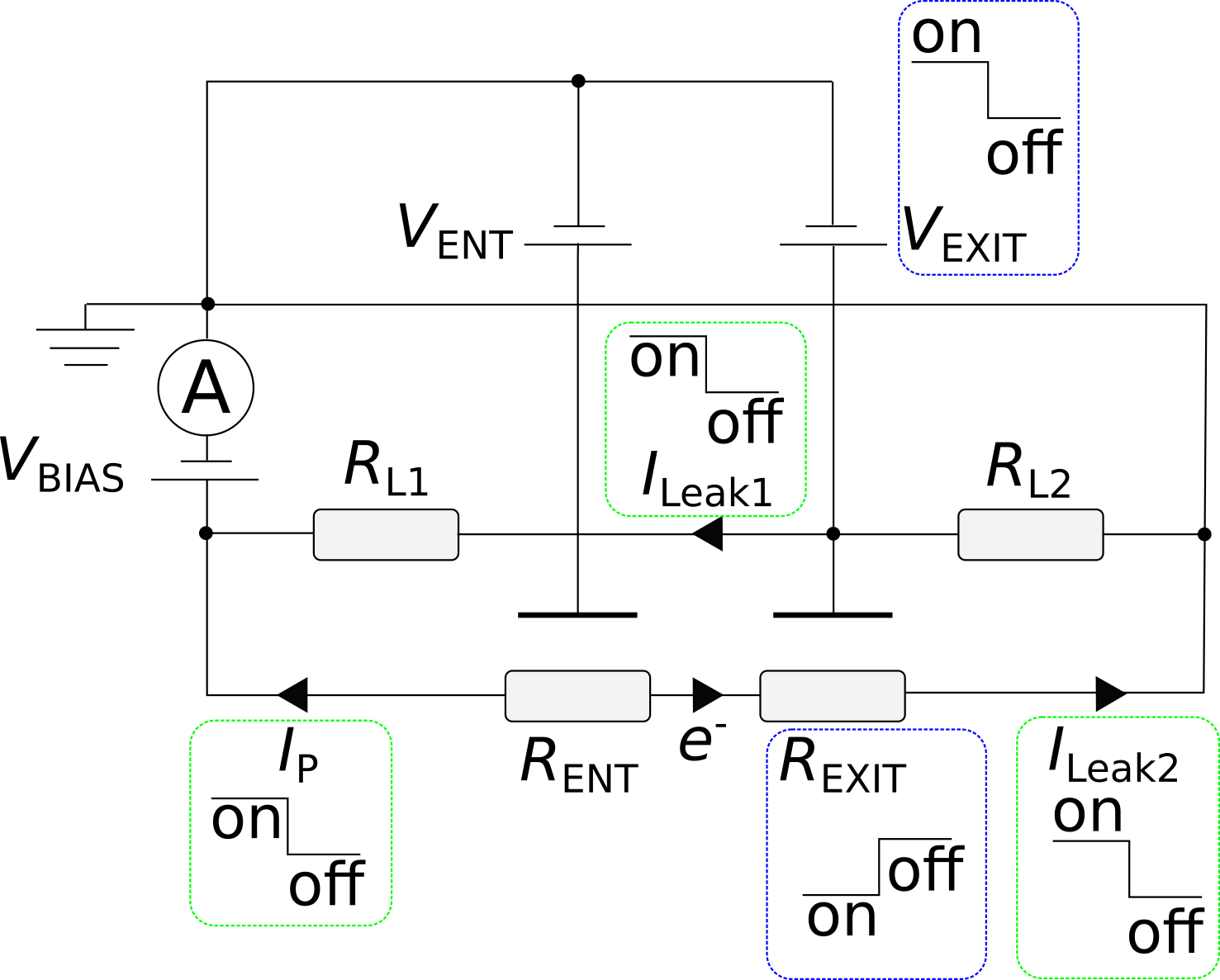}
\caption{\label{LeakagePathFig}\textsf{Schematic circuit diagram illustrating two possible sources of leakage current when the on-off cycle is implemented by stepping the exit gate voltage. The ULCA is depicted as an ammeter in series with a stray bias voltage source $V_{\text{BIAS}}$ at the left of the diagram. The potential barriers forming the pump are shown as large resistances $R_{\text{ENT}}$ and $R_{\text{EXIT}}$. The meanings of the other circuit elements are explained in the accompanying supplementary text. }}
\end{figure}

Precision measurements carried out using the gate-stepping on-off cycle are corrected for leakage currents which, as noted in the previous supplementary section, are driven through nominally insulating parts of the sample, sample holder and experimental wiring. These currents are present all the time in precision measurements, but are assumed constant when the gate voltages are constant. A schematic circuit diagram in figure \ref{LeakagePathFig} illustrates two possible sources of leakage current. Here, the sample has been reduced to an electrical circuit consisting of two large resistances in series, $R_{\text{ENT}}$ and $R_{\text{EXIT}}$, caused by the potential barriers under the entrance and exit gates. When the pump drive signal is on, the entrance gate voltage is oscillating, but for the purposes of this analysis $R_{\text{ENT}}$ can be considered the time-averaged value of the entrance barrier resistance. Electrons are pumped from left to right, and hence the ULCA connected on the source side of the device measures a pump current $I_{\text{P}}$ with a positive sign. A leakage current $I_{\text{Leak1}}$ is driven by $V_{\text{EXIT}}$ through resistance $R_{\text{L1}}$, which is the parallel sum of leakage resistances in the sample, sample holder, and experimental wiring. As an order-of-magnitude estimate, typical isolation between leads in a cryogenic wiring setup is $10^{15}$~$\Omega$. Thus, a gate voltage of $-1$~V drives a current of $1$~fA, or $6$~ppm of $I_{\text{P}}$ in our experiment. When $V_{\text{EXIT}}$ is stepped to a more negative value for the `off' part of the cycle, $I_{\text{Leak1}}$ becomes more negative. This yields an on-off leakage error current with the same sign as $I_{\text{P}}$: the current evaluated from the on-off cycle is larger than the pump current. Note that leakage current also flows in the resistance $R_{\text{L2}}$ to the drain lead of the pump, but as this current does not pass through the ammeter it does not cause an error.

An additional source of leakage current is the stray bias present at the ULCA input, $V_{\text{BIAS}}$. This drives a DC current $I_{\text{Leak2}}$ through the pump. The large potential barriers formed by the entrance and exit gates at the pump operation point, combined with the low stray bias of the ULCA make this current very small: for $V_{\text{BIAS}}=10$~$\mu$V and $R_{\text{ENT}}$, $R_{\text{EXIT}} \sim 10^{12}$~$\Omega$, $I_{\text{Leak2}} \sim10^{-17}$~A, which is less than $0.1$~ppm of $I_{\text{P}}$. When $V_{\text{EXIT}}$ is switched to a more negative value to turn the pump off, $I_{\text{Leak2}}$ decreases. This gives an on-off leakage error current which depends on the sign of $V_{\text{BIAS}}$, and which has the opposite sign to $I_{\text{P}}$ for the sign of $V_{\text{BIAS}}$ indicated in the diagram. 

Measurements made at MIKES and NPL using the gate-stepping on-off cycle, denoted NPL 2 and MIKES 2 in Fig. \ref{SummaryHighAccFig}, are corrected for leakage current: $I_{\text{P}} = \Delta I-\Delta I_{\text{Leak}}$. Here, $\Delta I$ is the on-off difference current from the precision pump current measurement with the AC pump drive turned on, and $\Delta I_{\text{Leak}}$ is the on-off difference current in a leakage measurement with the AC pump drive turned off. Measurements of $\Delta I_{\text{Leak}}$ at NPL and MIKES gave $91 \pm 67$~aA and $(-162 \pm 50)$~aA respectively. Note the different signs, consistent with the possibility outlined above of two different leakage mechanisms. After correcting for leakage currents, the measurements exploiting the gate-stepping cycle were consistent with those made using the power switching cycle.

\begin{figure}
\includegraphics[width=9cm]{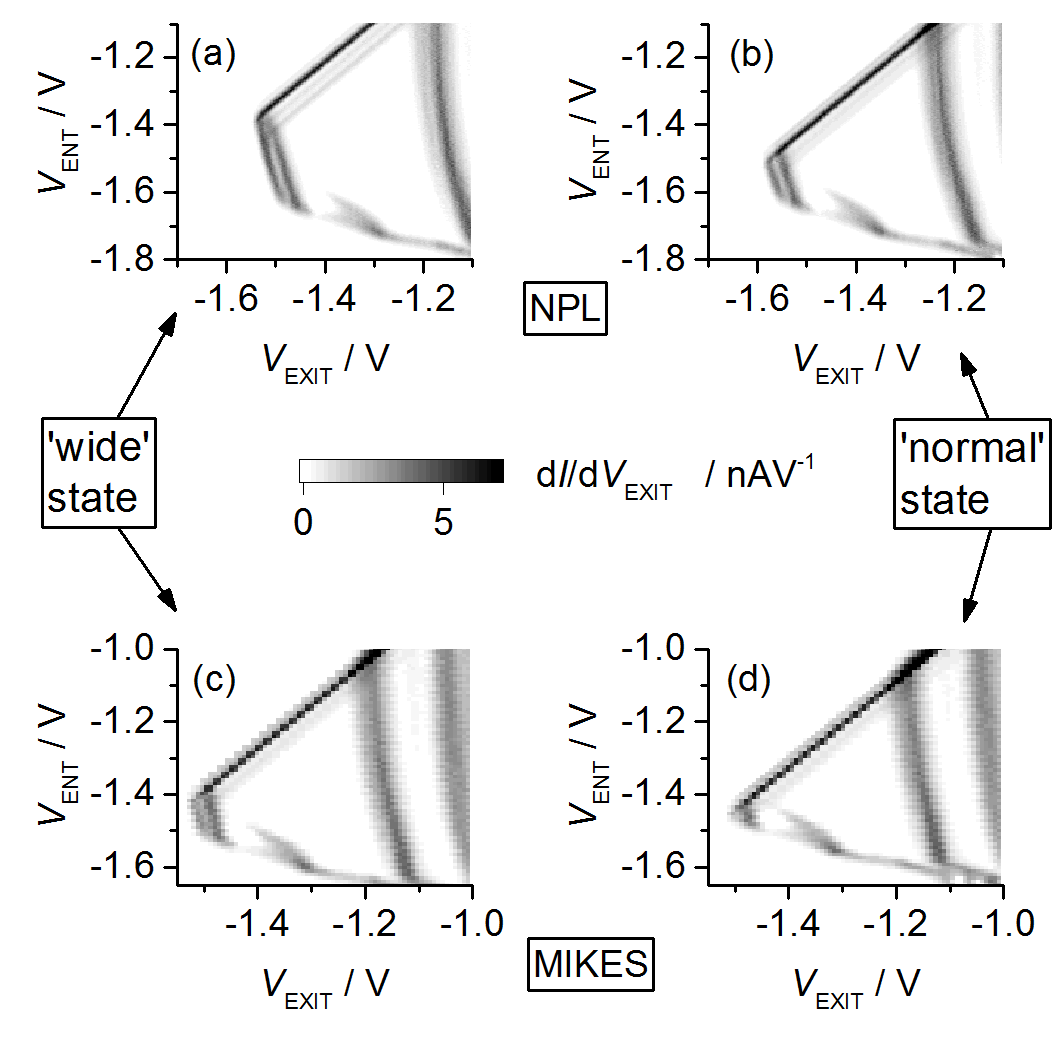}
\caption{\label{BistableFig}\textsf{Pump maps illustrating the `wide' (panels a and c) and `normal' (panels b and d) states exhibited by the pump. Panels (a) and (b) were measured at NPL with $P_{\text{RF}} = 11.6$~dBm, and panels (c) and (d) were measured at MIKES with $P_{\text{RF}} = 12.25$~dBm.}}
\end{figure}

\subsection{Bistability of pump map}

The pump map was usually stable for several weeks at liquid helium temperatures, but it occasionally switched to a different state characterised by a more extended plateau along the $V_{\text{ENT}}$ axis, as well as a transition to the $N=1$ plateau at slightly more positive $V_{\text{EXIT}}$. We refer to this state as the `wide' state, to distinguish it from the `normal' state. Transitions to the wide state did not occur with sufficient regularity to determine a cause, but at NPL they appeared to be correlated with setting the gate voltages to zero. Once in the wide state, the pump switched back to the normal state within 1-2 days. Example pump maps in figure \ref{BistableFig} show the normal and the wide states measured at NPL and MIKES. To maintain consistency, all the data presented in this paper was taken with the pump in the normal state. 

\subsection{Low-noise cryogenic wiring}
Simply lowering a device into a dewar of liquid helium is a very simple way of cooling it to a temperature of $4$~K. However, the dewar presents presents a demanding environment for low-noise current measurements. Bubbles can form in the liquid and cause vibration, while the temperature gradient along the wires can change as the liquid level drops, causing current noise spikes due to triboelectric processes as the wires are subject to changing mechanical strain. Many tests were carried out on different types of wiring to determine the best configuration for low-noise current measurements in a helium dewar. These tests, and their conclusions, will be the subject of a future paper. Here, we summarise the final experimental configuration. The DC experimental wiring consists of 10 enameled constantan wires of approximately $0.4$~mm diameter (not all of the wires are used in this experiment). They are stuck together into a flat ribbon, with no twisting, using general electrical varnish. It is important that the ribbon consisted of parallel wires, rather than the commonly-used set of twisted pairs. Twisting the wires introduces strain which can be a source of triboelectric current spikes, and is in any case not as important for current measurement as it is for voltage measurement, where the loop area must be minimised to prevent inductive coupling of interference. The loom is threaded through a $3$~mm inside-diameter stainless steel tube which runs from the breakout box at the top of the probe to approximately $5$~cm above the sample holder. The purpose of this tube is to contain the wiring loom and prevent it from vibrating within the larger (approximately 1 cm inside-diameter) structural tube of the probe. The `cold' ends of the wires are terminated with SSMB connectors to mate with the NPL-designed sample holder, and the `warm' ends terminate inside the breakout box which has BNC connections. The room-temperature cable from the probe breakout box to the ULCA input is the specialised low-noise cable supplied with the ULCA. A final important detail on the cryogenic probe construction is that the use of insulating tape is kept to a minimum throughout the probe, as insulating tape can carry charge which will generate a current in a wire nearby if that wire vibrates.

\end{document}